\documentclass[aps,prb,superscriptaddress,amsmath,amssymb,showpacs,twocolumn]{revtex4-1}

\usepackage{graphicx}
\usepackage[utf8]{inputenc}
\usepackage[T1]{fontenc}
\usepackage{mathtools}
\usepackage{hyperref}
\usepackage{units}
\usepackage{color}

\newcommand{\EP}{\varepsilon_P}
\newcommand{\EC}{\varepsilon_c}
\newcommand{\Ed}{\varepsilon_0}
\newcommand{\Edt}{{\tilde \varepsilon}_0}

\newcommand{\refe}[1]{Eq.~(\ref{#1})}
\newcommand{\reff}[1]{Fig.~\ref{#1}}

\newcommand{\autocorrelator}[1]{\left\langle #1 (t) #1 (0) \right\rangle}
\newcommand{\avg}[1]{\left\langle #1\right\rangle}

\newcommand{\rem}[1]{}
\newcommand{\beq}{\begin{equation}}
\newcommand{\eeq}{\end{equation}}

\begin{document}
\title{
Electro-Mechanical Transition in Quantum dots}

\author{G. Micchi}
\author{R. Avriller}
\author{F. Pistolesi}
\affiliation{
Univ. Bordeaux, LOMA, UMR 5798,  Talence, France.\\
CNRS, LOMA, UMR 5798, F-33400 Talence, France.\\
}

\begin{abstract}
The strong coupling between electronic transport in a single-level quantum dot and a 
capacitively coupled nano-mechanical oscillator may lead 
to a transition towards a mechanically-bistable and blocked-current state.
Its observation is at reach in carbon-nanotube state-of-art experiments. 
In a recent publication [Phys.\ Rev.\ Lett.\ {\bf 115}, 206802 (2015)] we have shown that this transition is 
characterized by pronounced signatures on the oscillator mechanical properties: 
the susceptibility, the displacement fluctuation spectrum and the ring-down time. 
These properties are extracted from transport measurements, however the relation between the 
mechanical quantities and the electronic signal is not always straightforward. 
Moreover the dependence of the same quantities on temperature, bias or gate voltage, and 
external dissipation has not been studied.
The purpose of this paper is to fill this gap and provide a detailed description of the transition.
Specifically we find: 
({\em i}\/)  The relation between the current-noise and the displacement spectrum. 
({\em ii}\/) The peculiar behavior of the gate-voltage dependence of these spectra at the transition.
({\em iii}\/) The robustness of the transition towards the effect of external fluctuations and dissipation.
\end{abstract}

\date{\today}


\maketitle


\section{Introduction}     


Detection and actuation of mechanical systems at the nano-scale is a present-day challenge
that has important fundamental and application 
perspectives.\cite{roukes_plenty_2007,mamin_nuclear_2007,lassagne_ultrasensitive_2008,rabl_quantum_2010,
mahboob_phonon-cavity_2012, mahboob_phonon_2013,moser_ultrasensitive_2013,aspelmeyer_cavity_2014} 
Electronic transport allows for detection of the displacement of extremely small mechanical oscillators, 
like carbon nanotubes.\cite{sazonova_tunable_2004}
Increasing the coupling between electronic transport and mechanical displacement allows for a more sensitive
detection and more precise actuation of the mechanical motion. 
However, strong coupling can also lead to an unexpected behavior of the system, as it was
first recognized in studying coherent electronic transport in molecular junctions coupled 
to low-frequency (classical) vibrations.\cite{galperin_hysteresis_2005,mozyrsky_intermittent_2006}
In nano-mechanical systems the importance of the electron back-action  
on the mechanical degrees of freedom was recognized very 
early,\cite{armour_classical_2004,blanter_single-electron_2004,chtchelkatchev_charge_2004,
clerk_quantum_2005,doiron_electrical_2006,usmani_strong_2007}
and concerned the case of oscillators coupled to single-electron transistors 
in the orthodox incoherent Coulomb blockade regime.
In this regime it was found that for sufficiently strong coupling the system presents a mechanical bistability with
a suppression of the current,\cite{pistolesi_current_2007} 
analogous to the coherent case of molecular junctions.\cite{pistolesi_self-consistent_2008} 
In both cases the transition to the bistable state is controlled by the intensity of the electro-mechanical 
coupling, $\epsilon_P$, defined as $F_0^2/k$, where $k$ is the spring constant of the oscillator and 
$F_0$ is the force acting on the oscillator when one electron is added to the quantum 
dot or to the metallic island of the single-electron transistor.
The bistability and suppression of the current is expected when both the temperature $T$ 
and the bias voltage $V$ are smaller than $\epsilon_P$ 
(we work throughout the paper with the notation that the electric charge $e$, 
the Boltzmann constant $k_B$, and the reduced Planck constant $\hbar$ are equal to 1). 
In the coherent transport case an additional condition has to be satisfied: $\epsilon_P$ larger than 
$\Gamma$, the typical width of the electron state.
Observation of this effect is not easy, since in general the value of $\epsilon_P$ is very small. 
In order to increase its value it has been proposed, for instance, to study a mechanical system close to the
Euler-buckling instability, since there $k$ vanishes, leading to an increase of two orders 
of magnitude of $\epsilon_P$.\cite{weick_discontinuous_2010,weick_euler_2011,bruggemann_large_2012}
However the transition has not been observed yet, contrary to the related, but different effect, known as 
Franck-Condon blockade.\cite{braig_vibrational_2003,koch_franck-condon_2005,koch_theory_2006,leturcq_franckcondon_2009}
The latter concerns high-frequency phonons in the regime $\Gamma \ll T \ll \omega_0$, where 
$\omega_0/2 \pi$ is the phonon frequency.

The experimental situation changed recently with the impressive progress that has been achieved in the 
detection and manipulation of carbon-nanotubes mechanical oscillators.
After the first observation of the mechanical resonance\cite{sazonova_tunable_2004}
it has been possible to observe the back-action of single-electron transport on the mechanical 
mode.\cite{lassagne_coupling_2009,steele_strong_2009} 
This opened the way to an impressive series of applications for 
mass,\cite{lassagne_ultrasensitive_2008,chaste_nanomechanical_2012}
force,\cite{moser_ultrasensitive_2013}, magnetic\cite{ganzhorn_carbon_2013} and biological sensors,\cite{kosaka_detection_2014}
high quality factors resonators,\cite{laird_high_2011,moser_nanotube_2014} as well as  the 
observation of phenomena related to the non-linear behavior of the oscillator.\cite{eichler_nonlinear_2011,zhang_interplay_2014}
In particular, recently the group of S. Ilani reported the observation of devices with values of $\epsilon_P$ of 
the order of 0.3 K.\cite{waissman_realization_2013,benyamini_real-space_2014}
The system they study is well modelled by a quantum dot coupled to a mechanical oscillator. 
In a very recent publication we investigated some of the features expected when the transition is studied 
varying the coupling constant $\epsilon_P$ at low temperature
($\omega_0 \ll V, T \ll \epsilon_P$).\cite{micchi_mechanical_2015}
Specifically we found that the mechanical mode softens at the transition to the bistable state, with a minimal value 
controlled by the bias voltage, and that phase fluctuations dominate the dynamics leading to a universal 
quality factor of 1.71.
This provided a first glance on the transition scenario, but several points deserve to be 
investigated in order to clarify the full picture.

In this paper we analyze the current-noise spectrum and, in particular, we investigate 
its relation to the displacement spectrum.
The two spectra may be proportional to each other, since the current is modulated by 
the displacement, but in general more complex relations exist.

One of the clearest proofs of the back-action of single-electron transport on the 
mechanical degrees of freedom is the observation of a weak softening of the mode as a function
of the gate voltage around the degenerate point. 
Since at the transition the mode softens spectacularly, we investigate how this reflects on the 
gate-voltage dependence of the mechanical resonance.

In general the system is out of equilibrium and the bias voltage plays the role of the temperature. 
Both the bias voltage and the temperature influence the transition, mainly smearing it.
We thus investigate the region where its observation is still possible.

Finally real systems are always coupled to other degrees of freedom, that induce dissipation.
We consider the effect of the presence of a finite dissipation. 
We find that it can either have no effect on the main physical quantities, or improve the visibility
of the transition when the temperature of the environment is lower than the bias voltage.

The plan of the paper is the following. 
In Sec.~\ref{sec:model} we introduce the microscopic model describing a single-level quantum dot coupled to a 
mechanical oscillator, we present then a Langevin and Fokker-Planck description of the system dynamics. 
In Sec.~\ref{Sec_Current_measurements} we derive the average electric current through the device.
In Sec.~\ref{Sec_current_fluctuations} we obtain and study the current-fluctuation spectrum relating it 
to the displacement spectrum. 
In Sec.~\ref{sec:gate_sweeping} we study the gate voltage dependence of the aforementioned spectra. 
In Sec.~\ref{sec:temperature} the effect of thermal and off-equilibrium fluctuations is discussed. 
In Sec.~\ref{sec:dissipation} the influence of an external damping is considered on the oscillator quality factor. 
Finally Sec.~\ref{sec:conclusions} gives our conclusions and perspectives.

%
%
%
%

\section{Model and Prerequisites}     
\label{sec:model}

\subsection{The model}

Following the literature on the subject\cite{mozyrsky_intermittent_2006,micchi_mechanical_2015} we 
describe the quantum dot by a spinless single electronic level coupled to a mechanical oscillator.
The Hamiltonian reads:
\begin{equation}
 	H = H_L + H_R + H_T + (\Ed-F_0 x) d^\dag d^{\phantom \dag} + \frac{p^2}{2m} + \frac{k x^2}{2} \,,
 	\label{eq:Hamiltonian}
\end{equation}
where $d$ and $\Ed$ are the destruction operator and the energy of the electronic level on the dot, respectively, 
$x$ is the displacement of the relevant mechanical mode,
$p$ its conjugated momentum,
$m$ the mode effective mass,
$k$ the spring constant (with $\omega_0=\sqrt{k/m}$),
and $F_0$ the electrostatic force acting on the oscillator when the electronic level is occupied. 
The first three terms describe the leads and their coupling to the electronic level:
$H_\alpha = \sum_k (\epsilon_{\alpha k} - \mu_\alpha)  c_{\alpha k}^\dag c_{\alpha k}^{\phantom \dag}$, 
$H_T = \sum_k t_{\alpha} c_{\alpha k}^\dag d^{\phantom \dag} + {\rm h.c.}$,
with $\alpha = L(R)$ for the left (right) lead, $c$ and $\epsilon_{\alpha k}$ the destruction operator and the energy 
of the electrons in the leads, respectively, and $\mu_\alpha$ the chemical potential.
From these quantities one can define the lead's tunneling rate
$\Gamma_\alpha \equiv \pi t^2_\alpha \rho_\alpha$ with $\rho_\alpha$ the density of states 
and the single-level width $\Gamma \equiv \Gamma_L+\Gamma_R$.
For simplicity in the following we choose $\Gamma_L = \Gamma_R = \Gamma/2$ and  $\mu_L =-\mu_R= V/2$, with 
$V$ the bias voltage.
In the following we will always consider the regime that is realized in current experiments with carbon nanotubes.
This corresponds to the limit $\Gamma_\alpha \gg \omega_0$ and $\omega_0 \ll \max(T, V)$ that 
allows one to solve the problem within the Born-Oppenheimer approximation and to treat 
the mechanical mode classically.\cite{mozyrsky_intermittent_2006} 
We will not consider in this paper the case of strongly coupled fast (quantum) oscillator, that is 
more relevant for molecular transport.\cite{koch_theory_2006,pistolesi_cooling_2009,
avriller_electron-phonon_2009,piovano_coherent_2011}


\subsection{Classical stochastic description}
\label{sec:langevin}

With these assumptions the displacement dynamics of the mechanical mode can be described by 
a Langevin equation:
\begin{equation}
 	m \ddot x + A(x) \dot x + m\omega_0^2 x = F_e(x) + \xi(t) \, ,
 \label{eq:langevin}
\end{equation}
where the charge fluctuations on the dot are at the origin of 
the average force $F_e(x)$, the dissipation 
$A(x)$, and the stochastic force $\xi(t)$, 
that satisfies $\langle \xi(t)\xi(t')\rangle=D(x) \delta(t-t')$ 
(actually the time range of the correlation function
is $\Gamma^{-1}$, but since $\Gamma\gg \omega_0$ 
we approximate it by a Dirac delta function).\cite{mozyrsky_intermittent_2006, pistolesi_current_2007}
The coefficients read $F_e(x) =  F_0 n(x)$, 
$ D(x) = F_0^2 S_{nn}(x, \omega=0)$, 
$A(x) = -F_0^2 \left.(\partial S_{nn}/ \partial \omega)(x, \omega)\right|_{\omega=0}$, 
with $n=d^\dag d$, $n(x) = \langle n \rangle$ the average occupation of the dot, 
and $S_{nn}(x, \omega) = \autocorrelator{n}_\omega$ the charge fluctuation spectrum.
The explicit expression for $A$, $F_e$, and $D$ has been obtained in Ref.~\onlinecite{mozyrsky_intermittent_2006}
(we provide an alternative and more elementary derivation in App.~\ref{app:langevin}):
\begin{align}
 \langle n\rangle = &\int_{-\infty}^{+\infty} \frac{d\omega}{2\pi\Gamma} \ \left(f_L + f_R\right) \tau
  \label{eq:FAD_electronicN}
 \\
 \left. S_{nn} \right|_{\omega=0}  = &\sum_{\alpha, \beta} \int_{-\infty}^{+\infty} \frac{d\omega}{2\pi\Gamma^2} f_\alpha (1 - f_\beta) \tau^2 
 \label{eq:FAD_electronicD}
 \\
 \left. {dS_{nn} \over  d\omega} \right|_{\omega=0}= &\sum_{\alpha, \beta} \int_{-\infty}^{+\infty} \frac{d\omega}{2\pi\Gamma^2} f_\alpha \tau \left[f_\beta'\tau - (1 - f_\beta) \tau'\right],
 \label{eq:FAD_electronicA}
 \end{align}
where $f_\alpha(T, \omega) = f_F(\omega-\mu_\alpha)$, $f_F(\omega)=(1+e^{\omega/T})^{-1}$ is the Fermi function, 
$\tau(\omega, x) = 1/\{1 + [(\omega-\Ed + F_0 x)/\Gamma]^2\}$ is the energy-dependent
electronic transmission factor through the quantum dot, 
and the prime in \refe{eq:FAD_electronicA} indicates the derivative with respect to $\omega$.

Since the stochastic force is short-ranged in time one can derive a Fokker-Planck equation for the probability distribution 
$P(x,p,t)$:\cite{blanter_single-electron_2004,blanter_erratum:_2005}
\begin{equation}
 \partial_t P = \frac{p}{m} \partial_x P - F \partial_p P + \frac{A}{m} \partial_p (pP) + \frac{D}{2} \partial_p^2 P \,,
 \label{eq:FP}
\end{equation}
where $F(x)=-kx+F_e(x)$.


\subsection{Effective potential determined by $F(x)$}
\label{sec:mean_field}

\begin{figure}
 \includegraphics[width = \linewidth]{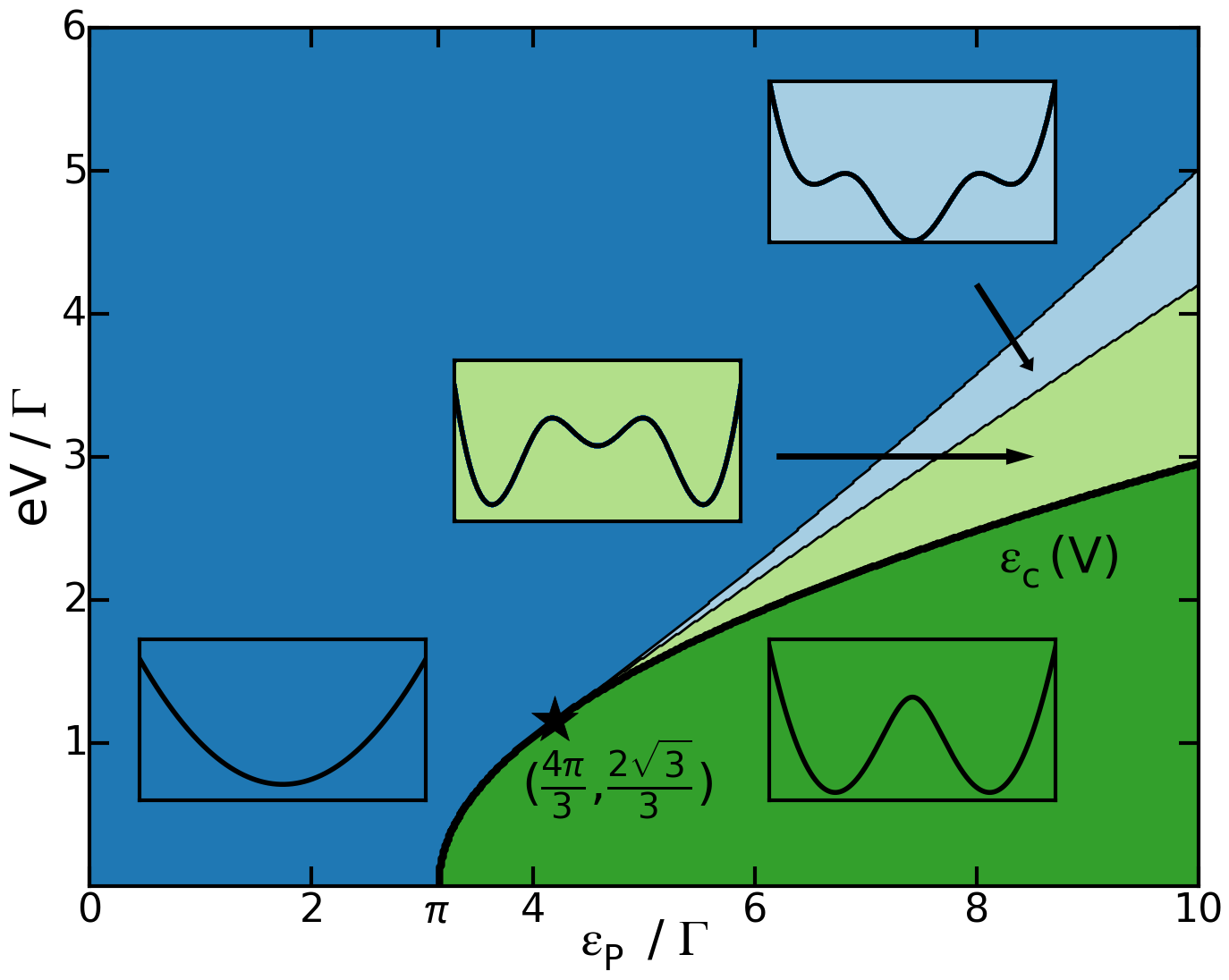}
 \caption{
 (Color online) Phase-diagram for the stability of the effective potential in the $V-\EP$ plane for
 $\Ed=\EP/2$ and vanishing temperature.
 The thick line indicates the critical line $\EC$.
 }
 \label{fig:phase_diagram}
\end{figure}

The total deterministic force $F(x)$ acting on the oscillator is given by the sum of the mechanical 
restoring force $-k x$ and of the electronic contribution $F_0 \langle n  \rangle$.
For vanishing temperature (the finite temperature case will be discussed in Sec.~\ref{sec:temperature})
and arbitrary bias voltage one finds:
\begin{equation}
	F(x) 
	= 
	-kx + \frac{F_0}{2} + \frac{F_0}{2\pi} \sum_{\alpha = \pm} \arctan \frac{\alpha V/2 - \Ed +F_0 x}{\Gamma} \,.
 \label{eq:force}
\end{equation}
One can verify that for $\Ed= \EP/2$ the force is anti-symmetric with respect to the point 
$x_0= F_0/2k$.
This value of the gating is special and corresponds to the new electron-hole symmetry point 
for the system.
We will focus mainly on this case in the remainder of the paper, dedicating 
Sec.~\ref{sec:gate_sweeping} to a discussion on the  $\Ed$-dependence of 
different physical quantities.
The force is more conveniently expressed in terms of the dimensionless variable $y=F_0(x-x_0)/\Gamma$,
for which one can write a Taylor expansion in $y$:
\begin{equation}
	 F(y) = F_0 \sum_{n=0}^\infty a_{2n+1} y^{2n+1}, 
	 \label{ExpF}
 \end{equation}
with 
\beq
	a_1 = -\Gamma/\EP+ \arctan'(V/2\Gamma)/\pi
\eeq
and $a_n= \arctan^{(n)}(V/2\Gamma)/[\pi n! ]$ for $n>1$.
From the form of $F$ it follows immediately the form of the effective potential defined 
as  
\beq
	U(x)=-\int^x F(x')dx'. \label{Ueff}
\eeq
As a function of $y$ it reads $U(y)= -\Gamma \sum_{n=0}^\infty {a_{2n+1} }  y^{2n+2}/(2n+1)$
The condition $a_1<0$ determines the stability of the minimum in $y=0$ and gives the equation
$\EP < \EC(V)$ with
\beq
	\EC(V) = \pi \Gamma \left(1+{V^2\over 4 \Gamma^2} \right)
	\label{criticalV}
\eeq
defining the critical line.
By graphical solution of the equation $F(y)=0$ one finds that there is always 
one solution at $y=0$, and then there can be one or two additional pairs of solutions 
at $\pm y_1$ and $\pm y_2$. 
Since the function is regular, the stationary points corresponds either to maxima or minima of 
the potential, that alternate. 
Moreover the system has always at least one stable solution, thus when 
the solution at $y=0$ is unstable there can be only two stable solutions
at $\pm y_1$ (3 solutions).
%
This proves that in the region of the phase diagram $\EP > \EC(V)$ the 
potential has two symmetric minima (see also Fig.~\ref{fig:phase_diagram}).
The situation is more complex when the origin is stable ($a_1<0$), but $a_3$ and $a_5$ 
are not both negative. 
In this case the origin can be either the only minimum or two additional side minima 
may appear. 
By direct calculation one finds that  $a_3>0$ for $V>2\Gamma/\sqrt{3}$.
Thus along the critical line this point ($\EC(2\Gamma/\sqrt{3})=4\pi/3\Gamma$)
marks the beginning of the multi-stability. 
The full behavior is shown in Fig.~\ref{fig:phase_diagram}, as obtained by numerically solving
$F(y)=0$. 

%
%

In the following we will limit to voltages smaller than
$2\Gamma/\sqrt{3}$, thus the two possible phases are the stable and bistable one,
separated by the line $\EC(V)$.
For vanishing voltage and temperature the effective potential reads:
\begin{equation}
 	U(y) = \frac{\Gamma}{2\EP} y^2  -
  			\frac{\Gamma}{2\pi} \left[2y\arctan y - \log\left(1+y^2\right)\right],
  			\label{UzeroT}
\end{equation}
that expanded around $y=0$ gives:
\begin{equation}
U(y)= \frac{\Gamma}{2\pi} \frac{\pi\Gamma}{\EP} \left[\left(1-\frac{\EP}{\pi\Gamma}\right) y^2  + \frac{1}{6} \frac{\EP}{\pi\Gamma} y^4
+O(y^6)
\right].
 \label{eq:U_approx}
\end{equation}
In this case the transition takes place at the value $\EP = \pi \Gamma = \EC(0) \equiv \EC$.

The first important consequence of the transition is the softening of the mechanical mode. 
Let's define 
\beq
	\omega_m^2=-{1\over m} {dF\over dx} 
	\label{defomegam}	
\eeq 
for $x$ at the minimum of the potential.
This gives 
\beq
	\omega_m^2=\omega_0^2(1-\frac{\EP}{\EC})
	\label{omegaMono}
\eeq
for $\EP \leq \EC$ and 
\beq
	\omega_m^2 \approx 2\omega_0^2(\frac{\EP}{\EC}-1)
	\label{omegaBi}
\eeq
for $\EP \geq \EC$ and $(\EP-\EC)/\EC \ll1$.
As discussed in Ref.~\onlinecite{micchi_mechanical_2015} the softening of the mechanical mode has important consequences
that can be observed in the response functions. 

\begin{figure}
\includegraphics[width=\linewidth]{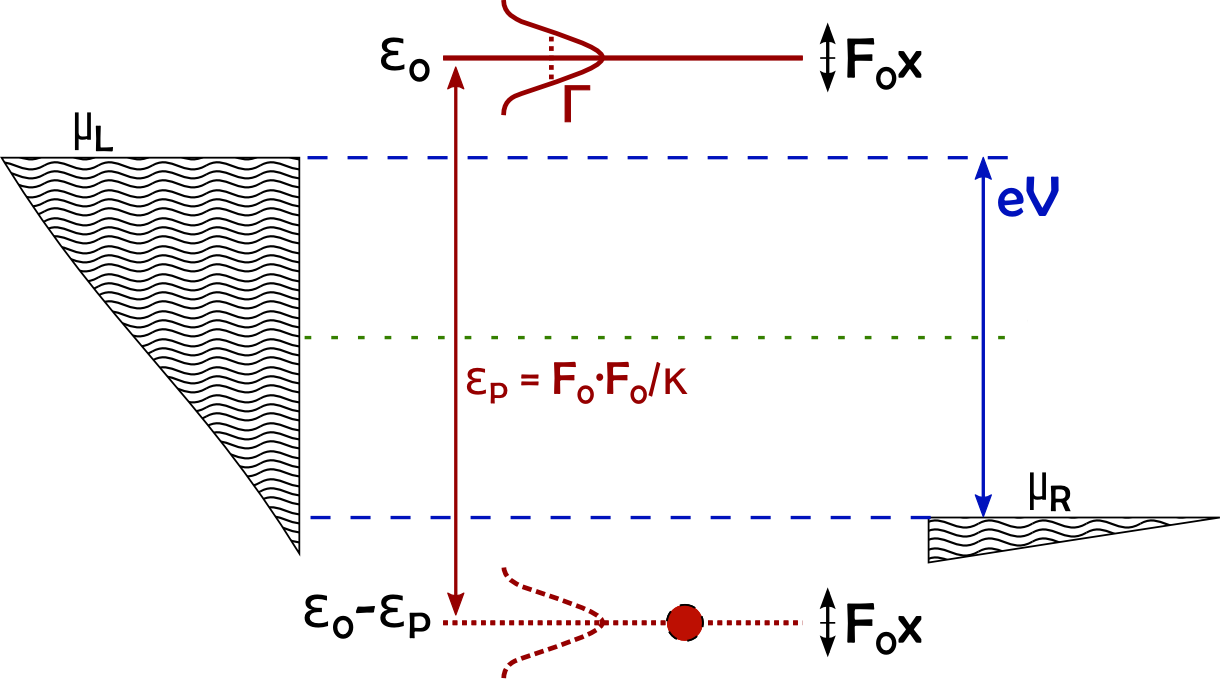}
\caption{(Color online)
A pictorial view of the current blockade mechanism. 
We show the position of the electronic level of the dot for the two stable positions of the 
oscillators depending on the state of the dot, empty (upper level) or full (lower level).
When $\EP>V$  both levels are out of the conducting window
and for thus $\Gamma \ll V, \EP $ the current cannot flow stabilizing 
both states. 
For $\Gamma$ of the order of $V$ and $\EP$ the picture is blurred by the finite width of the 
energy level, leading simply to a reduction of the current.
}
\label{fig:scheme}
\end{figure}

The appearance of the bistability marks also the beginning of the reduction of the current, as we will discuss 
more in details in the following. 
In Fig.~\ref{fig:scheme} we provide a simple picture of the blockade:
The presence of an electron changes also the
stable position of the oscillator and consequently the energy of its level ($\epsilon_0$).
When the shift in the energy ($\EP$) exceeds the bias voltage both the occupied and empty state are 
out of the bias window, thus in a blocked current state. 
This holds strictly when $\Gamma \ll \EP$, but in general a residual current is possible, due to the finite width in
energy of the single electronic level.

%
%

\subsection{Stationary solution for $V,T \ll \Gamma$}
\label{sec:FP}

In the regime $V,T \ll \Gamma$ the stationary solution of the Fokker-Planck equation (\ref{eq:FP}) 
can be found analytically.
As a matter of fact the coefficients $A$ and $D$ depend weakly on $x$.
In general when their ratio is independent of the position it is possible 
to define an effective temperature:
\beq
	T_\text{eff} = \frac{D(x)}{2A(x)} \,.
\eeq
One can verify by substitution that the function
\beq
	P_\text{st}(x, p) = {\mathcal N}^{-1} e^{-E(x, p)/T_\text{eff}},
	\label{GibbsP}
\eeq
solves \refe{eq:FP}.
Here $E(x,p)=p^2/2m+U(x)$ is the oscillator energy, $U$ is defined by \refe{Ueff}, and 
$\mathcal N = \int dx \, dp \, \exp\left(-\frac{E(x, p)}{T_\text{eff}}\right)$
a normalization factor. 
The explicit expressions for $A$ and $D$ in the limit  $V,T \ll \Gamma$ can be calculated by 
exploiting the weak $\omega$-dependence of $\tau(\omega, x)$ with respect to the fast variation of the
Fermi functions. 
Expanding the expression of $\tau(\omega,x)$ around $\omega=0$ and integrating in $\omega$ we obtains:
\beq
A(x) = \frac{F_0^2}{\pi\Gamma^2} \tau_0^2(x) \left[1 + \frac{2\pi^2}{3} \left[5\tau_0(x)  - 6\tau_0^2(x)\right] \left(\frac{T}{\Gamma}\right)^2 \right],
\eeq
where $\tau_0(x)=\tau(0,x)$, with the remarkable relation
\beq
	\frac{D(x)}{2A(x)}= {T\over 2} \left(1 + \frac{V}{2T} \coth \frac{V}{2T} + \dots \right) \equiv 	T_\text{eff}
	\label{DefTeff}
\,.
\eeq
We thus find that at lowest order in $T/\Gamma$ and $V/\Gamma$ the ratio $D/A$ does not depend on $x$.
This is not surprising for $V\ll T$, since in this case the system is essentially in equilibrium and 
$T_\text{eff}\approx T$, but it is somewhat unexpected for $(\Gamma \gg) V \gg T$,
{\em i.e.}\ far from equilibrium.
In this limit we find $T_\text{eff}\approx V/4$ that agrees with what found for a metallic single-electron transistor
in Ref.~\onlinecite{armour_classical_2004}.
Most numerical simulations presented in this paper are performed for $V/\Gamma=5\cdot 10^{-3}$, thus in this regime. 

From the form of the stationary solution it is clear that the effective potential plays a crucial role.
It is thus convenient to introduce a parameterization of the phase space ($x$-$p$) of the oscillator in terms of the
energy $E(x,p)$ and of the time along a trajectory of given energy:
\beq
	 t_E(x, p) = {1\over (2m)^{1/2}} \int^x \frac{dx'}{\left(E(x',p)-U(x') \right)^{1/2} }.
\eeq 
The Jacobian of the transformation is unitary.
In terms of the new variables the distribution is uniform along the trajectory, thus leading to the following form for 
the stationary distribution integrated over the trajectories
\beq
	P_\text{st}(E) = {2 \pi \over \omega(E)} e^{-E/T_\text{eff}}/{\cal N}
\eeq
with ${\cal N}=\int dE 2 \pi  e^{-E/T_\text{eff}}/ \omega(E)  $, and $2\pi/\omega(E)$ 
the period of the trajectory of energy $E$.

Lets finally briefly discuss $\omega(E)$. 
This can be obtained analytically for the potential containing only the quadratic and quartic part.\cite{dykman_fluctuations_1980,dykman_time_1980}
We give its expression in the stable phase $\EP<\EC$.
\begin{equation}
 \frac{\omega(E)}{\omega_0} =
 \frac{\pi}{2} \frac{\sqrt{B(E)}}{K[-m(E)]}\,,
\end{equation}
with
\begin{equation*}
 \begin{gathered}
  B(E) = \frac{\EC-\EP}{2\EC} + \sqrt{\frac{3 (\EC-\EP)^2 \EC + 4\pi^2 \EP^2 E}{12 \EC^3}}\,,\\
  m(E) = \frac{\sqrt{(\EC-\EP)^2 + 4\pi^2 \EP^2 E/3\EC} - (\EC-\EP)}{\sqrt{(\EC-\EP)^2 + 4\pi^2 \EP^2 E/3\EC} + (\EC-\EP)}
 \end{gathered}
\end{equation*}
and $K[-m(E)]$ the complete elliptic integral of the first kind with parameter $-m(E)$.
At the critical point $\EP = \EC$ the expression simplifies:
\begin{equation}
 {\omega(E) \over \omega_0} = \left(\frac{\pi^3}{48}\right)^\frac{1}{4} \frac{\Gamma[3/4]}{\Gamma[5/4]} \, \left({E}/{\Gamma}\right)^\frac{1}{4} \approx 1.212 \left({E}/{\Gamma}\right)^\frac{1}{4}.
 \label{omegaE}
\end{equation}
In the other limit, for $\EP \ll \EC$ one can expand in $E$ obtaining
\beq
	\omega(E) = \omega_m + \omega' E
	\label{omegamExp}
\eeq
with
\beq
\omega'
\equiv 
\frac{d\omega_m}{dE} =
 \frac{\pi^2\omega_0 }{4\EC}  \left(\frac{\varepsilon_P}{\varepsilon_c}\right )^2 
 \left(1 - \frac{\varepsilon_P}{\varepsilon_c} \right)^{-3/2} \, .
\label{domegamdE}
\eeq

In the next sections we will consider in some details how different physical quantities behave as a
function of the coupling strength $\EP$, particularly at the transition to the bistable phase.

%
%

\section{Electronic current}

\label{Sec_Current_measurements}

\begin{figure}
 \includegraphics[width=\linewidth]{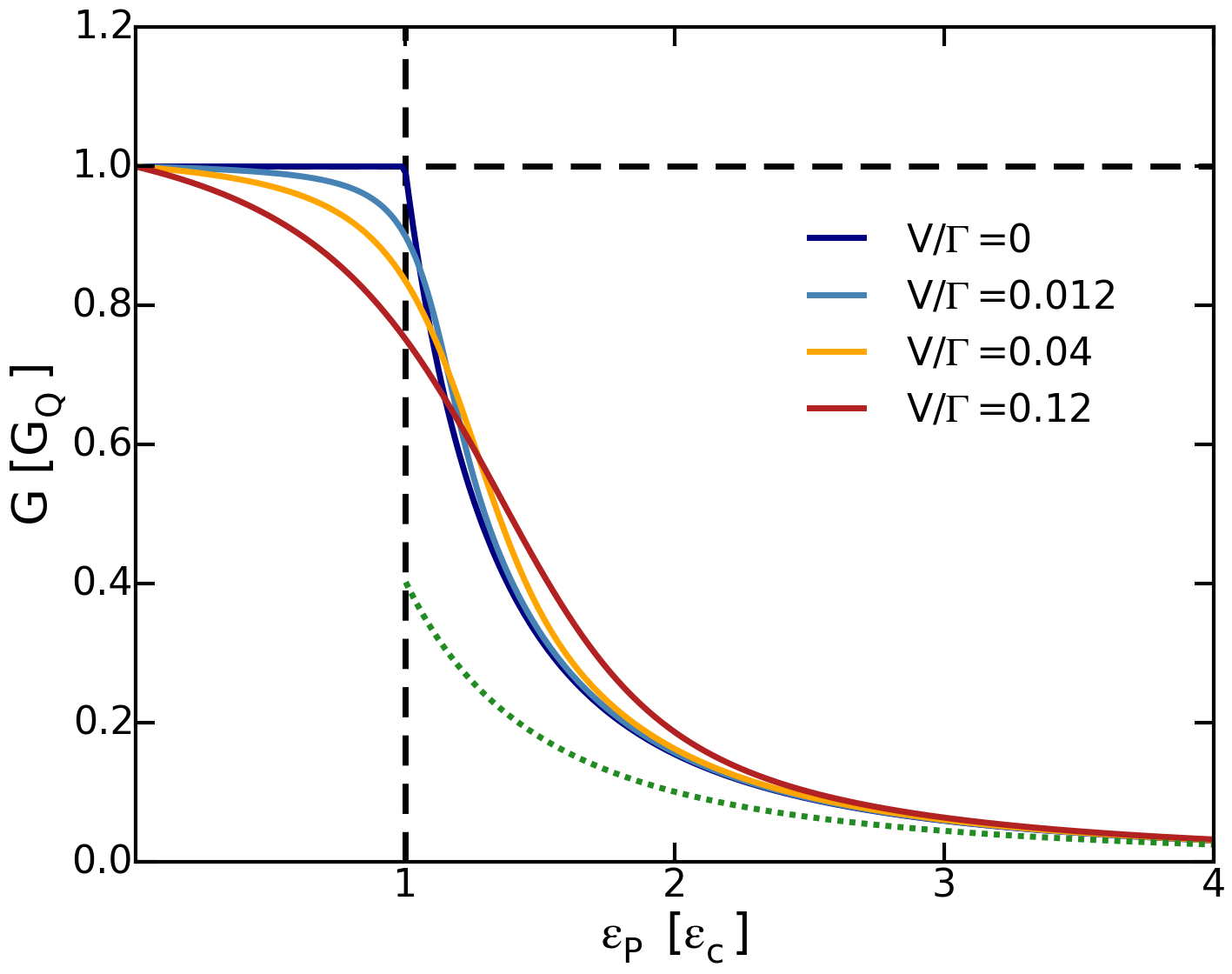}
 \caption{(Color online) 
 Conductance as a function of the coupling constant $\EP$ for $T=0$
and different values of $V$ as indicated in the legend. 
 The dotted (green) line is the asymptotic behavior of the conductance obtained
 for $\EP \gg \EC$, when the system is in the bistable phase. 
 }
 \label{fig:conductance}
\end{figure}

We begin by considering the current flowing through the quantum dot. 
In the classical regime considered in this paper for given value of $x$ one can write:
\beq
	I(x)= \int {d\omega\over 2\pi}  \tau(\omega,x) \left[f_L(\omega)-f_R(\omega)\right]
	\label{Ieqx}
		\,.
\eeq
The current is the average of this quantity over the positions visited by the oscillator:
\beq
	I(t)= \int \frac{d\omega}{2\pi} \int dxdp  P(x,p,t) I(x)
	\label{Ieq}
		\,.
\eeq
In the stationary regime $I$ does not depend on $t$ and one can use the stationary distribution for 
the probability. 
We define the  low voltage conductance as $G(V)=I/V$ for small $V$. 
For $T=0$ and $V\ll \Gamma$ it reads:
\beq
G(V)= G_Q \int \!\! dx dp  P_\text{st}(x,p) \tau(0,x) + O(V/\Gamma)
\label{EqGV}
\,,
\eeq
where $G_Q=1/2\pi$ is the quantum of conductance in the units $\hbar=e=1$.
Fig.~\ref{fig:conductance} shows the conductance evaluated numerically from 
\refe{EqGV} for different values of $V$, $T=0$, and $\Ed=\EP/2$.
The extreme limit $V\rightarrow 0$ shows very well the presence of a transition for 
$\EP=\EC$, the conductance is flat for $\EP<\EC$ and decreases smoothly for 
$\EP>\EC$, with a cusp at $\EP=\EC$.
Already at this values it is clear that the suppression of the current is relatively slow as a function of 
the coupling. 
This is due to the Lorentzian decay of the transparency $\tau$ as a function of the energy. 
For higher values of the voltage the transition is smooth. 
%
%
%
%
%

When the effective potential is truncated to fourth order [cf. \refe{eq:U_approx}] it is possible to 
obtain an analytic expression of the conductance.
The dot-transparency can be expressed as a power series of $y$: 
$\tau(0,x)= \sum_{n=0}^{+\infty}\left(-1 \right)^n y^{2n}$.
By expanding the $y^4$ term in the exponential of the stationary distribution one can 
evaluate all the Gaussian integrals for $\EP \leq \EC$ obtaining for the conductance:
\begin{equation}
 \left\langle G \right\rangle = G_Q 
 \sum_{n=0}^{+\infty}\left(-1 \right)^n 
 \left(\frac{12 \pi^2 T_\text{eff} }{\EC} \right)^{n/2}
\Xi_n\left\lbrack \alpha\right\rbrack
\label{eq:I_general}
\end{equation}
with
\beq
	\Xi_n\left\lbrack \alpha\right\rbrack = 
	\frac{
	\sum_{p=0}^{+\infty} \frac{(-\alpha )^p}{p!}
	\mathbf{\Gamma}\left\lbrack \frac{1+2(n+p)}{4}
	\right\rbrack
	}{
	\sum_{p=0}^{+\infty} \frac{\left(-\alpha \right)^p}{p!}
	\mathbf{\Gamma}\left\lbrack \frac{1+2p}{4}
	\right\rbrack
	}
	\,,
\eeq
$\alpha = {(\EC-\EP)}/{\EP}\sqrt{3\EC/{\pi^2}T_{\rm eff}}$, and 
$\mathbf{\Gamma}\left\lbrack x \right\rbrack$ the Euler-Gamma function.

The expression simplifies for $\EP\ll \EC$: 
\begin{equation}
G = G_Q \left(1 - {\pi^2\over 4} {\EP V \over \EC^2 }\right)
\,. 
\label{eq:I_weak}
\end{equation}
Close to the transition (specifically for $\alpha \ll 1$) we find
\begin{equation}
 \begin{aligned}
  G(V) & \approx G_Q 
  \Big\{   1 - \left( \frac{3\pi^2V }{\EC} \right)^{\frac{1}{2}} \frac{\mathbf{\Gamma}[3/4]}{\mathbf{\Gamma}[1/4]}+\\
  & - 6 \frac{\EC-\EP}{\EP} \frac{\mathbf{\Gamma}[3/4]}{\mathbf{\Gamma}[1/4]}
  \left[
  \frac{\mathbf{\Gamma}[3/4]}{\mathbf{\Gamma}[1/4]} -
  \frac{\mathbf{\Gamma}[5/4]}{\mathbf{\Gamma}[3/4]}
  \right]
  \Big\}
  \label{eq:I_crit}\, .
 \end{aligned}
\end{equation}
This indicates that $G(V)$ decreases rapidly with $V$ at criticality, as can be seen in the figure.

Finally we investigate the behavior deep in the bistable regime.
For $\EP\gg\EC$ the two stable solutions are at $y=\pm \EP/2\Gamma$. 
By simply taking the value of the transparency for these values of $y$ we have
\beq
	G={G_Q \over 1+(\pi \EP/2\EC)^2} \,.
	\label{GG}
\eeq
The conductance vanishes as a power law for large coupling.
We show this curve dotted on Fig.~\ref{fig:conductance}.
This value can be corrected by taking into account the Gaussian fluctuations 
around the two minima that add a correction $1+3\EC V/2\EP^2$
that increases a bit the conductance by increasing the effective temperature.

In conclusion we have seen that the current is slowly suppressed as a function of the 
coupling constant, but in general there is not a spectacular effect at the transition. 
This is in contrast to what happens for large values of the coupling and finite voltage $V\gg\Gamma$, for 
which the current has quite a sharp dependence on the voltage.\cite{pistolesi_self-consistent_2008}
In the following we will consider the current-noise spectrum.
We will see that it is directly related to the mechanical dynamics and presents thus clear signatures
of the transition.

%
%
%
%

\section{Power spectral density of current fluctuations}
\label{Sec_current_fluctuations}

The interest in the measurement of the current-fluctuation spectrum is that it allows one to detect the motion of the
mechanical oscillator without driving it.
The current is modulated by a variation of the displacement, thus each fluctuation of the displacement will 
appear also as a fluctuation of the current. 
In Ref. \onlinecite{micchi_mechanical_2015} we have investigated in some details the behavior of the displacement
fluctuation spectrum [$S_{xx}(\omega)$] at the transition.
As it can be seen from Eqs.~(\ref{omegaMono}) and (\ref{omegaBi}) the mechanical mode becomes soft at the 
transition as predicted by the effective potential behavior.
At the transition the fluctuations are very strong, and the meaning of mechanical mode is not so clear anymore. 
That is why it is necessary to consider measurable manifestations of the mechanical dynamics. 
The study of the displacement fluctuation spectrum allows to follow the evolution of the mechanical mode 
and to see its behavior at the transition.
We refer to Ref.~\onlinecite{micchi_mechanical_2015} for a detailed discussion, but it may be useful to recall the main 
features: 
The position of the main peak of the spectrum has a minimum as a function of the 
coupling constant at $\EP=\EC$, the minimum value is not zero, but a finite value that is controlled by 
the bias voltage (or the effective temperature).
It vanishes like $(T/\Gamma)^{1/4}$ and the quality factor is universal with value 1.7. 
The displacement spectrum is related to the response function to an external weak driving force by a relation of 
the fluctuation-dissipation kind.
A direct measurement can be possible by detection of the current noise.
But the relation between the displacement fluctuation spectrum and the current fluctuation spectrum is not 
always straightforward, in particular close to the transition.
In this section, we discuss the behavior of the current-correlation function and its relation to the
displacement fluctuation spectrum.

We begin with its definition:
\begin{equation}
	S_{II}(\omega) = {1\over 2} \int\displaylimits_{-\infty}^{+\infty} dt e^{i\omega t}
	\langle \tilde I(t) \tilde I(0)+I(0) \tilde I(t) \rangle
	\label{eq:SII} \, ,
\end{equation}
where $\tilde I(t)=I(t)-\langle I \rangle$ is the quantum current operator 
(note that we define the spectrum with a factor of 1/2 with respect to the usual convention for 
quantum noise).
The time-scale separation between the fast electronic degrees of freedom and the slow mechanical ones 
allows to distinguish the purely electronic contribution to the current noise, coming from the granularity 
of the charge, and the one coming from the slow fluctuations of the position of the oscillator.

Let's first discuss briefly the electronic contribution. 
For fixed and given $x$ the electronic noise reads 
$S^{e}_{II}(\omega \ll \Gamma)=\tau(x)[1-\tau(x)] \langle I \rangle$.\cite{blanter_shot_2000}
The spectrum is flat for $\omega \ll \Gamma$, and to obtain the observed value 
it is sufficient to average the above expression with $P(x,p)$. 
In the following we will neglect this contribution, since it has no frequency dependence at the interesting 
mechanical frequencies.

We thus consider only the current fluctuations originating from the displacement fluctuation.
We can then write 
\begin{equation}
	S_{II}(\omega) = \int\displaylimits_{-\infty}^{+\infty} dt e^{i\omega t}
	\langle \tilde I[x(t)] \tilde I[x(0)] \rangle
	\label{eq:SIIxx} \, ,
\end{equation}
with $I(x)$ given by \refe{Ieqx}. 
Starting from this expression one can compute numerically the average current following the method presented 
in Ref.~\onlinecite{pistolesi_self-consistent_2008}.
It implies the numerical solution of the Laplace transform of the Fokker-Planck equation and 
it gives
\begin{equation}
 S_{II}(\omega)= -2 {\rm Tr} \left[(\hat I - \left\langle I\right\rangle)
 \frac{\mathcal L_0}{\omega^2 + \mathcal L_0^2} (\hat I - \left\langle I\right\rangle) P_\text{st} \right],
 \label{eq:Sii_pistolesi}
\end{equation}
where $\hat I$ is the operator that at each point in the $x$--$p$ phase space associates
the current \refe{Ieqx}, $\mathcal L_0$ is the Fokker-Planck matrix that comes from writing
 \refe{eq:FP} as $\partial_t P = \mathcal L_0 P$, 
 and $P_\text{st}$ is the steady-state solution satisfying $\mathcal L_0 P_\text{st}=0$.

 We will use this numerical method in the following, but before it is instructive to study the explicit relation between
 the current spectral density and the displacement spectrum defined as:
 \beq
 	S_{xx}(\omega) 
 	= 
 	\int_{-\infty}^{+\infty}{e^{i\omega t}\left\langle \tilde x(t)  \tilde x(0)\right\rangle dt},
\eeq
with $\tilde x(t)=x(t)-\langle x \rangle$.
Since we are mainly concerned with the case $V,T \ll \Gamma$ we can linearize the current in the voltage.
The current correlation function can then be written as:
\begin{equation}
 S_{II}(t) = G^2_Q V^2 
 \left\langle  \tilde \tau\lbrack x(t)\rbrack \tilde \tau\lbrack x(0)\rbrack\right\rangle  \, ,
 \label{eq:SII_2}
\end{equation}
where  $\tilde \tau=\tau-\langle \tau \rangle$.
This expression shows that the relation between $S_{II}$ and $S_{xx}$ depends on the 
regime considered. 
When the relevant dependence of $\tau$ on $x$ is linear the result may be a direct proportionality, but 
in general more complex relation exist.

Finally the current correlator can be calculated analytically in the regime of very small damping rate 
$A /m \ll \Delta \omega$, with $\Delta \omega$ the width of the main peak in the spectral density 
induced by the oscillator non-linearity.\cite{micchi_mechanical_2015}
Following the method introduced by Dykman {\em et al.},\cite{dykman_time_1980} and Ref. \onlinecite{micchi_mechanical_2015}
the auto-correlation function for the current fluctuations in this regime can be well approximated by
\begin{equation}
 {S_{II}(t)  \over \left( G_Q V \right)^2} 
 = 
 \int\displaylimits_0^{+\infty} dE \int\displaylimits_{0}^{2\pi/\omega(E)} d\tau P_{st}(E, \tau)
 \tilde\tau \lbrack x_E(t+\tau)\rbrack
 \tilde\tau \lbrack x_E(\tau)\rbrack  \, ,
 \label{eq:SII_3}
\end{equation}
where $x_E(t)$ is the function with period $2\pi/\omega(E)$ that satisfies the equation of 
motion $m \ddot x=F(x)$ for given energy $E=p^2/2m+U(x)$.
In the next subsections we present the results for the current spectrum in different regimes. 


\subsection{$S_{II}$ in the mono-stable phase $\EP \leq \EC$}

\label{Sec_S_II_Weak-coupling}

For $\EP \leq \EC$ (and $V,T\ll \Gamma$) the effective potential has a single minimum at $x= F_0/2k$,
around which the system oscillates.
When the amplitude of these oscillations is sufficiently small
(essentially for $T_\text{eff}\ll \Gamma$, see also the section on the finite temperature) 
we can expand the transmission factor to second order in $x$:
\beq
	\tau \lbrack x(t) \rbrack \approx 1 - (F_0 x(t) / \Gamma)^2 \,.
\eeq
This leads to the following expression for the current fluctuations
\begin{equation}
 S_{II}(t) \approx \left( G_Q V \right)^2 \left( \frac{F_0}{\Gamma} \right)^4 
 \left\langle {\tilde x}^2(t) {\tilde x}^2(0)\right\rangle\, .
 \label{eq:SIIPerturbative_1}
\end{equation}
This relation allows one to relate the current spectrum to the displacement spectrum only in the 
weak coupling regime where the position fluctuations are still Gaussian. 
In this case using Wick theorem we obtain:
\begin{equation}
 S_{II}(t) \approx 2 \left( G_Q V \right)^2 \left( \frac{F_0}{\Gamma} \right)^4 S_{xx}^2(t)\, ,
 \label{eq:SIIPerturbative_2}
\end{equation}
or equivalently its Fourier transform
\begin{equation}
 S_{II}(\omega) = 2(G_Q V)^2 \left( \frac{F_0}{\Gamma} \right)^4
 \int\displaylimits_{-\infty}^{+\infty}\frac{d\omega'}{2\pi} S_{xx}(\omega-\omega') S_{xx}(\omega')\, .
 \label{eq:SIIPerturbative_3}
\end{equation}

\begin{figure}
 \includegraphics[width=\linewidth]{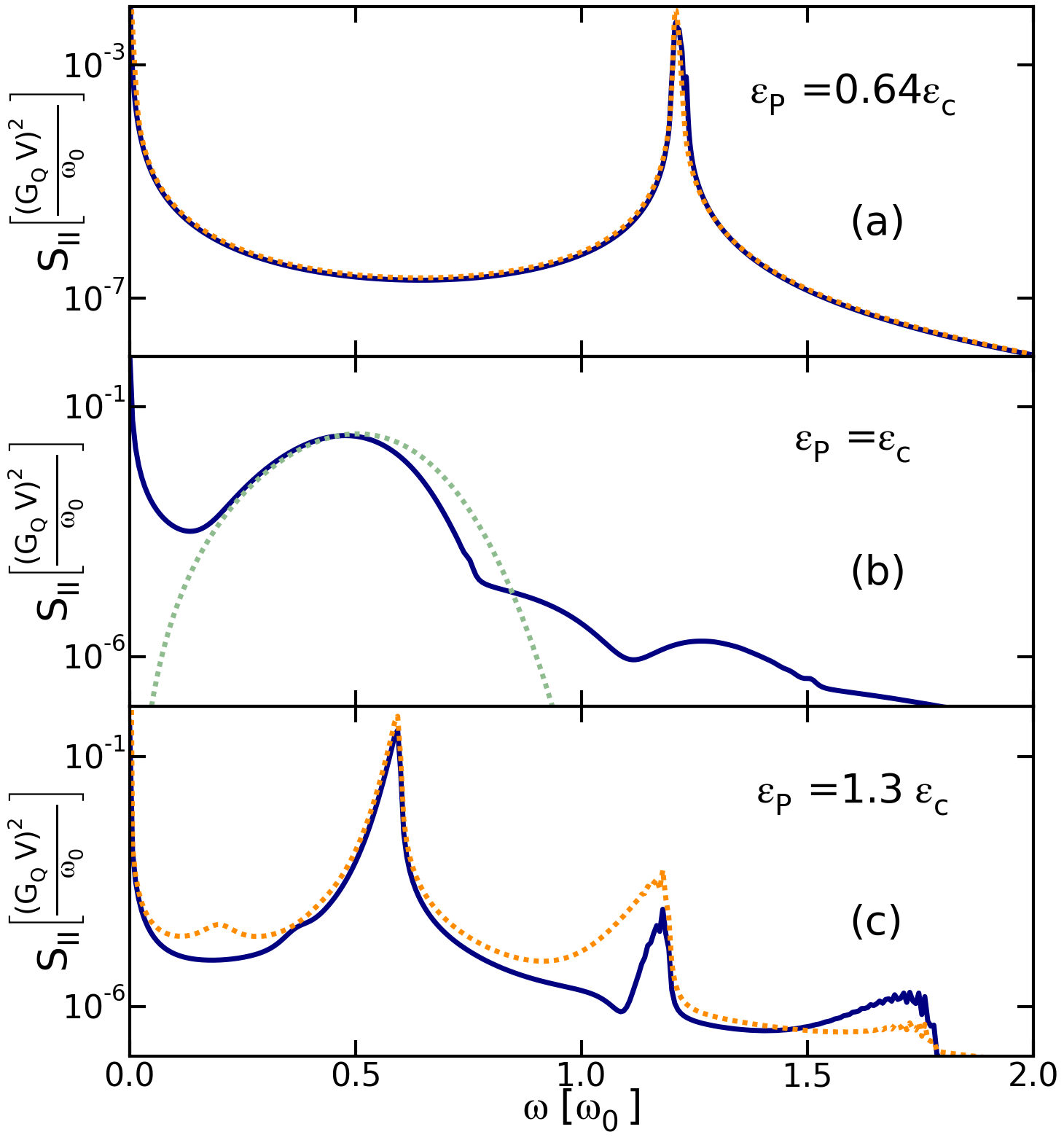}
\caption{(Color online)
 Power spectral density of current fluctuations $S_{II}(\omega)$ for different
 values of the coupling constant $\EP/\EC=0.64$, 1, 1.3 (as indicated in each panel)
 and for $\Gamma/\omega_0=10^3$, $V/\Gamma=5\cdot 10^{-3}$, $T=0$.
Blue solid lines give the result of the numerical evaluation of \refe{eq:Sii_pistolesi},
while dotted lines show analytical results:
In panel (a) the (orange) dotted line represents the convolution of the mechanical fluctuations $S_{xx}(\omega)$, as given in \refe{eq:SIIPerturbative_3}.
In panel (b) the (green) dotted line gives the analytical result of current fluctuations coming from \refe{eq:SII_regular}.
In panel (c) the (orange) dotted line is proportional to $S_{xx}$, as written in \refe{eq:SIIStrongCoupling_2}.
 }
 \label{fig:noise_ii}
\end{figure}

For extremely small $\EP \ll \EC$ the displacement spectrum $S_{xx}(\omega)$ shows two Lorentzian peaks
at frequencies $\pm\omega_m$ whose width is dominated by the dissipation $A/m\gg \Delta \omega$.
In this regime $S_{II}(\omega)$ will thus shows three Lorentzian peaks, centered at $\omega=0$ and 
$\omega=\pm 2\omega_m$ with a width of $2A/m$.
As discussed in Ref.~\onlinecite{micchi_mechanical_2015} the non-linear terms in the potential generate a finite 
width in the displacement spectrum $\Delta \omega$, and when $\Delta \omega\gg A/m$   
(for $\EP\ll\EC$ the condition reads $(V/\omega_0)(\EP/\EC) \gg 1$)
the Gaussian approximation cannot be used anymore to evaluate $S_{II}$.
We can nevertheless directly calculate the current spectrum using  \refe{eq:SII_3}.
We restrict to the (dominant) first harmonics  ($n = \pm 1$) in the Fourier series
$x_E(t)=\sum_{n=-\infty}^{+\infty}e^{-in\omega(E)t}x_n(E)$.
Expanding $\omega(E)$ in $E$ one finds that the spectral density has two contributions:
\beq
  S_{II}(\omega) \approx  S_{II}^\text{reg}(\omega) + S_{II}^\text{sing}(\omega)
  \label{eq:SIIPerturbative_4_tot}
\eeq
a regular ($S_{II}^\text{reg}(\omega)$) and a singular ($S_{II}^\text{sing}(\omega)$) one.
They read
\begin{eqnarray}
  {S_{II}^\text{reg}(\omega) \over (G_QV)^2 }&=&
  \frac{\pi}{4} \left(\frac{\EP}{\Gamma}\right)^2 \frac{\omega_m T_\text{eff}}{|\omega'|\Gamma^2 \omega} 
  f_1\left(\frac{\omega - 2\omega_m}{2\omega'T_\text{eff}}\right)
  \\
  {S_{II}^\text{sing}(\omega) \over (G_QV)^2 } &=&
  {7\pi^5 \over 2} 
{\EP^2\over \EC^4} {\omega_0^4 \over \omega_m^4} T_\text{eff}^2
  \delta(\omega)
 \end{eqnarray}
with $f_1(u) = u^2 e^{-u}$.
The singular contribution in reality is broadened by the dissipative term that if included in the calculation 
would give a width $2A/m$.
Note that this sharp low frequency contribution is not related to telegraph noise, as is the case in the 
bistable phase, since there is a single minimum of the potential. 
The location $\omega_M$ of the maximum of the power spectral density and the full width at half maximum $\Delta \omega$ of the spectral line read
\beq
  \omega_M = 2 \left(\omega_m + 2\omega'T_\text{eff}\right)
\eeq
with a width 
$  \Delta \omega \approx 6.79 \ \omega' T_\text{eff}$.
We also notice that the width is roughly twice the width found for the displacement spectrum 
and the position of the maximum is exactly at twice the position of the maximum of the displacement
spectrum. 
The numerical results obtained from \refe{eq:Sii_pistolesi} in this regime 
are shown in Fig.~\ref{fig:noise_ii} panel (a).
A comparison with the analytical calculation (dotted line) shows that the agreement is 
very good.
We observe that the peak at vanishing frequency appears also in the non-Gaussian limit
and that its width is of the same order of width dominated by the non-linear fluctuations 
of the peak at $\omega\approx 2\omega_m$.

%
%
%
%
%

%
%

%

We cannot extend the expansion too close to the critical point $\EP=\EC$, 
but exactly at the critical point the potential is purely quartic, $U(y) = ({\Gamma}/{12\pi}) y^4$,
and thus we can evaluate \refe{eq:SII_3} analytically.
The first harmonics of the oscillator displacement reads
$x_1(E)=\alpha_1 \Gamma/F_0 \left( 12\pi E/\Gamma \right)^{1/4}$
with a numerical factor $\alpha_1 \approx -0.477$ (see supplemental material of Ref.~\onlinecite{micchi_mechanical_2015}).
The spectrum has the form given by \refe{eq:SIIPerturbative_4_tot} with
\begin{align}
  S_{II}^\text{reg}(\omega) &\approx B_1 \frac{\left(G_QV \right)^{2}}{\omega_0}
  \left( \frac{T_\text{eff}}{\Gamma} \right)^{3/4} f_2\left(\frac{\omega}{\omega_M}\right)
  \label{eq:SII_regular}\\
  S_{II}^\text{sing}(\omega) &\approx B_2(G_Q V)^2\frac{T_\text{eff}}{\Gamma} \delta(\omega)\, ,
 \label{eq:SII_singular}
\end{align}
with a universal line-shape of the resonance given by $f_2(u)=u^6 e^{-3/2(u^4-1)}$ 
and 
$B_1=\alpha_1^4 96 \mathbf{\Gamma}(5/4)\pi^{5/4} 3^{7/4} (2e)^{-3/2}/\mathbf{\Gamma}(3/4)^2
\approx 6.797$, 
$B_2=\alpha_1^4 24 \pi^2 [4 \mathbf{\Gamma}(7/4) -\mathbf{\Gamma}(5/4)^2/\mathbf{\Gamma}(3/4)]/
\mathbf{\Gamma}(3/4)\approx 30.2$.

From these expression we infer that the maximum position and its width are 
\begin{equation}
  \omega_M \approx B_3 \,\omega_0\left(\frac{T_\text{eff}}{\Gamma}\right)^{1/4} \qquad
  \Delta \omega \approx 0.479 \,\omega_M
  \, ,
\label{Spread_S_II}
\end{equation}
with $B_3=\pi^{3/4} \mathbf{\Gamma}(3/4) 2^{-1/4}/\mathbf{\Gamma}(5/4)\approx 2.68 $.
In analogy to what found in Ref.~\onlinecite{micchi_mechanical_2015} for the width of the 
displacement spectrum, we find that the peak in the noise spectrum has a universal quality factor 
of $Q=(\omega_M/\Delta \omega)\approx 2.09$.
We note however that this differs from the one obtained for the displacement spectrum that is 1.71.
The difference is a signature of the non-Gaussian fluctuations, that break the simple relation 
(\ref{eq:SIIPerturbative_3}).
In Fig.~\ref{fig:noise_ii}-(b) we compare the prediction of \refe{eq:SII_regular} (green-dotted line)
with the full numerical calculation obtained with \refe{eq:Sii_pistolesi} (full line).
The overall shape of the main peak is well reproduced. 
We verified that the small over-estimation (the plot is in logarithmic scale) of the peak width 
is due to the absence in the analytical calculation of the six-order term of the potential.


\subsection{$S_{II}$ in the bistable phase $\EP>\EC$}
\label{Sec_S_II_Strong-coupling}

An analytical description is difficult close to the transition, but as soon as the potential minima 
become  sufficiently deep, we can again describe the oscillator 
as a harmonic oscillator around the new minima.
For $(\EP \gg \EC)$ thus the probability distribution becomes Gaussian around each minimum, related to 
the blocked current state in the occupied and empty electronic state. 
As a consequence, Wick theorem still holds if we expand displacement fluctuations 
separately around each minimum $y=\pm \pi \EP/2\EC$.
We can thus evaluate \refe{eq:SII_2} obtaining:
\begin{equation}
 \begin{aligned}
  S_{II}(t) &= 2 \left(G_Q V\right)^2
  \tau \left[ (1-\tau) \frac{F_0^2 S_{xx}(t)}{\Gamma^2} \label{eq:SIIStrongCoupling_1} +\right. \\
  &\, \tau^3 (3-4\tau)^2 \left(\frac{F_0^2 S_{xx}(t)}{\Gamma^2}\right)^2 + \\
  &\,  \left. 12\tau(1-\tau)(1-2\tau)\frac{F_0^2 \avg{x^2}}{\Gamma^2} \frac{F_0^2 S_{xx}(t)}{\Gamma^2}
  \right] \, .
 \end{aligned}
\end{equation}
(Note that we recover \refe{eq:SIIPerturbative_2} for the weak-coupling regime if we set $\tau = 1$.)
In the strong-coupling regime, $\tau \approx \left( 2\EC /\pi \EP \right)^2 \ll 1$ and the dominating term in \refe{eq:SIIStrongCoupling_1} is
\begin{eqnarray}
S_{II}(\omega) \approx 2\left( G_Q V \frac{F_0}{\Gamma} \right)^2
\tau (1-\tau)S_{xx}(\omega)
\label{eq:SIIStrongCoupling_2}\, .
\end{eqnarray}
Deep in the bistable regime we thus find that a simple linear relation between 
$S_{II}(\omega)$ and $S_{xx}(\omega)$ holds.
In Fig.~\ref{fig:noise_ii}-(c) we plot $S_{II}(\omega)$ obtained from \refe{eq:SIIStrongCoupling_2} 
in orange-dotted line, compared to the full numerical calculation of \refe{eq:Sii_pistolesi}.
We see that all the features are well reproduced with a reasonably good quantitative agreement.

%
%
%
%
%
%
%

\section{Gate voltage dependence}
\label{sec:gate_sweeping}

One of the most spectacular and direct proof of the back-action of the electronic transport on the 
mechanical dynamics is the observation of a maximum in the softening of the mechanical mode 
as a function of the gate voltage in coincidence with the maximum of the conductance of the quantum 
dot.\cite{lassagne_coupling_2009,steele_strong_2009,benyamini_real-space_2014}
As we have seen, the softening of the mechanical mode is also a signature of the transition to the 
bistable phase, but so far we have discussed only its dependence on the coupling $\EP$ when the 
gate voltage is adjusted so that  $\Ed=\EP/2$. 
In experiments the values of both $\EP$ and $\Ed$ are actually controlled by the value of the gate voltage.
The first corresponds essentially to the number of excess electrons on the gate lead, while the second 
is linearly related to the gate voltage. 
Close to the symmetry point $\Ed=\EP/2$ one can neglect the weak gate voltage dependence of $\EP$. 
It is then very interesting to study the behavior of the mechanical mode for small variation of the gate
voltage close to the symmetry point for given value of $\EP$. 
As for the previous case it is understood that the mode cannot be characterized only by the value of its 
resonance, since the spectrum has typically a complex behavior.
For this reason in this section we will study the displacement and current spectrum as a function of the gate 
voltage, or, as it appears in the present model, as a function of $\Delta \Ed=\Ed-\EP/2$.

\subsection{Softening of the mechanical mode from the effective potential}

Let's begin from the behavior of $\omega_m$ as defined by \refe{defomegam}.
For $V=T=0$ we can write the force as a function of $y$ as follows:
\beq
	{F(y)\over F_0}= -{\Gamma\over \EP} y + {1\over \pi} \arctan{(y-\Edt) }
	\label{Fyeq}
	\,.
\eeq
where $\Edt=\Delta \Ed/\Gamma$.
The presence of $\Edt$ changes the position of the solutions of $F(y)=0$, 
but by graphical analysis one can verify that there can be only either 1 or 3 solutions. 
Applying the definition of $\omega_m$ by calculating the first derivative of $F(y)$  at
$y_0$  one finds:
\beq
	{ \omega_m^2\over \omega_0^2 }
	=
	1-\tau(y_0) {\EP \over \EC}
	\label{omegamtau}
\eeq
where $\tau(y_0)=[1+(y_0-\Edt)^2]^{-1}$ is the transparency of the junction at $y_0$  
solution of the equation $F(y_0)=0$. 
The equation for $y_0$ can be written as
\beq
	y_0={\EP \over \EC} \arctan(y_0-\Edt) \,.
	\label{y0eq}
\eeq
\refe{omegamtau} generalizes \refe{omegaMono}. 
It is exact, but finding the value of $y_0$ may require some approximations. 
We consider thus three limiting cases. 

({\em i }) In the weak coupling limit for $\EP \ll \EC$ the value of $y_0$ is very small. 
Iterating \refe{y0eq} one can then write 
$y_0\approx {\EP \over \EC} \arctan({\EP \over \EC} \arctan(-\Edt)-\Edt)$ which has the correct 
small and large $\Edt$ behavior.
Keeping only the leading order one finds for the softening a Lorentzian behavior
\beq
	{ \omega_m^2\over \omega_0^2 }=1-{(\EP/\EC) \over 1+\Edt^2} 	,
\eeq
as observed in experiments.\cite{steele_strong_2009,lassagne_coupling_2009,benyamini_real-space_2014,waissman_realization_2013}

({\em ii}) At criticality for $\EP=\EC$ \refe{y0eq} can be solved for $\Edt \ll 1$ by expanding the right hand side 
to third order.
This gives $y_0=\Edt-(3 \Edt)^{1/3}$ and consequently 
$\omega_m=\omega_0 (3|\Edt|)^{1/3}$
that is singular at the transition.

Finally ({\em iii}) for  $\EP > \EC$ and $\Edt=0$ there exists two (stable) solutions of \refe{y0eq}
$y_0(\Edt=0)=\pm z_0$, with $z_0>0$.
%
For small value of $\Edt>0$ one can find the solution of  \refe{y0eq} as:
\beq
	y_0=-z_0+{(\Edt/4)\over \EP/\EC-1}
\eeq
that is valid for $0<\Edt \ll (\EP/\EC-1)^{3/2}$.
For $\Edt<0$ the stable solution is $-y_0$ [$y_0(\Edt)=-y(-\Edt)$], 
and sweeping $\Edt$ through 0 the system jumps from one solution to the other.
We can thus simply concentrate on the positive values of $\Edt$.
The linear dependence of $y_0$ leads to a linear dependence of $\omega_m^2$ close to $\Edt>0$.
For $0 < \EP-\EC \ll \EC$ one finds:
\beq
	{\omega_m\over \omega_0}
	=
	[2 (\EP/\EC-1)]^{1/2} + {\Edt \EP \over 2(\EP-\EC)} \left({3 \EC\over2 \EP}\right)^{1/2}
	\,.
        \label{eq:cusp}
\eeq
The slope diverges at $\EP\rightarrow \EC$, in agreement with the results at criticality that gives
$\omega_m \Edt^{1/3}$.
Since the curve is symmetric, the linear dependence leads to a cusp in the softening dependence on the 
gate voltage. 
This can be an indication of the bistability.

%
%
%
%
%
%
\begin{figure}
 \includegraphics[width = \linewidth]{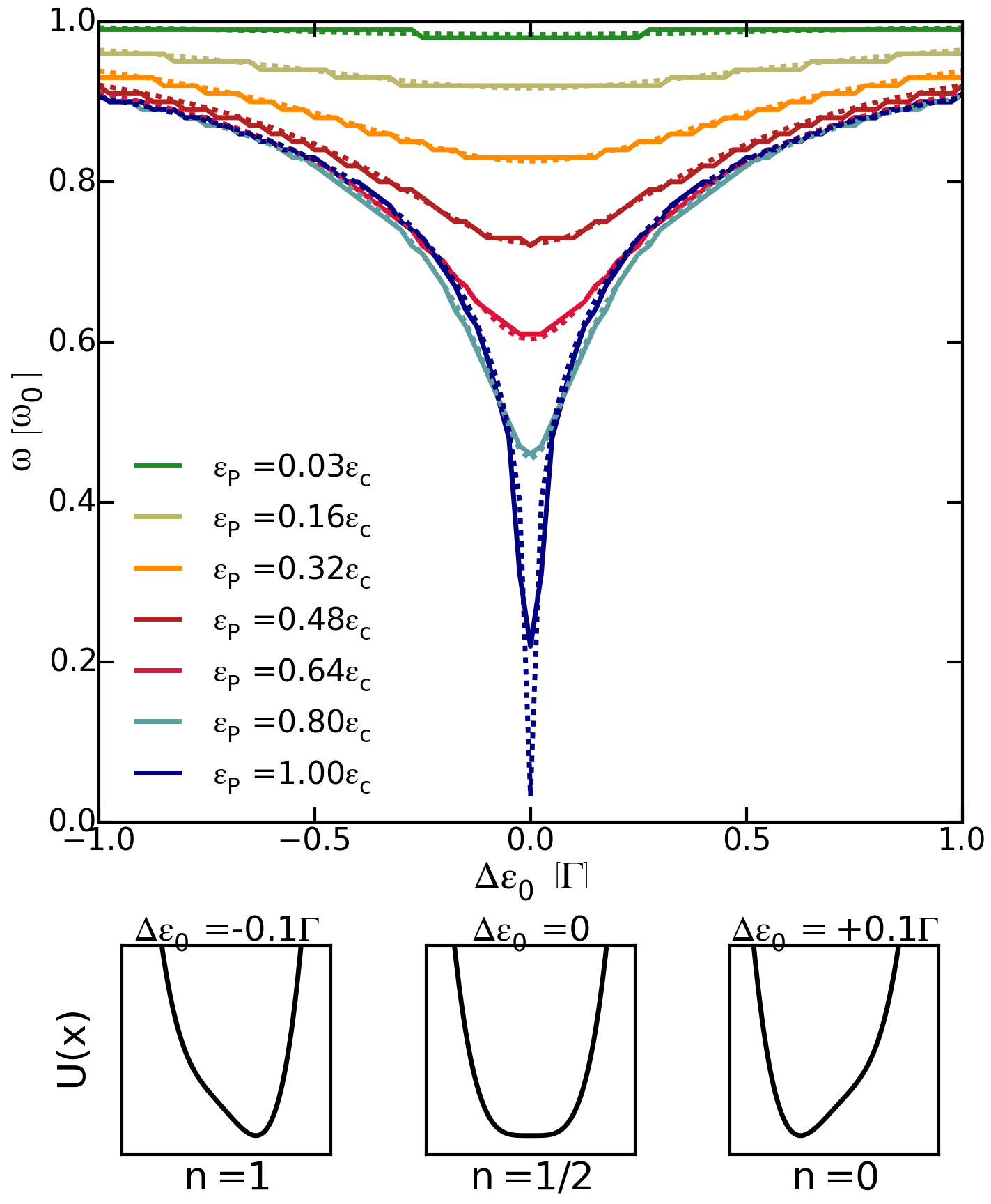}
 \caption{(Color online)
 Gate voltage dependence ($\Edt$) of the mechanical resonating frequency for different values 
 of the coupling $\EP\leq\EC$, as indicated in the legend. 
 The dotted lines indicate the numerical solution of the mean field  \refe{y0eq} and 
 \refe{omegamtau}.
 The continuous lines indicate the positions of the maxima in the $S_{xx}(\omega)$ spectrum.
 The parameters are $\Gamma/\omega_0 = 10^3$, $V/\Gamma = 5 \cdot 10^{-3} $, $T = 0$.
 Lower panel: Sketch of the effective potential as a function of the oscillator's position for various values of the dot's detuning;
 $n$ represents the occupation of the dot, and the vertical axis has a range of $4T_\text{eff}$.}
 \label{figVg1}
\end{figure}
%
%
%
%
%
%
\begin{figure}
 \includegraphics[width = \linewidth]{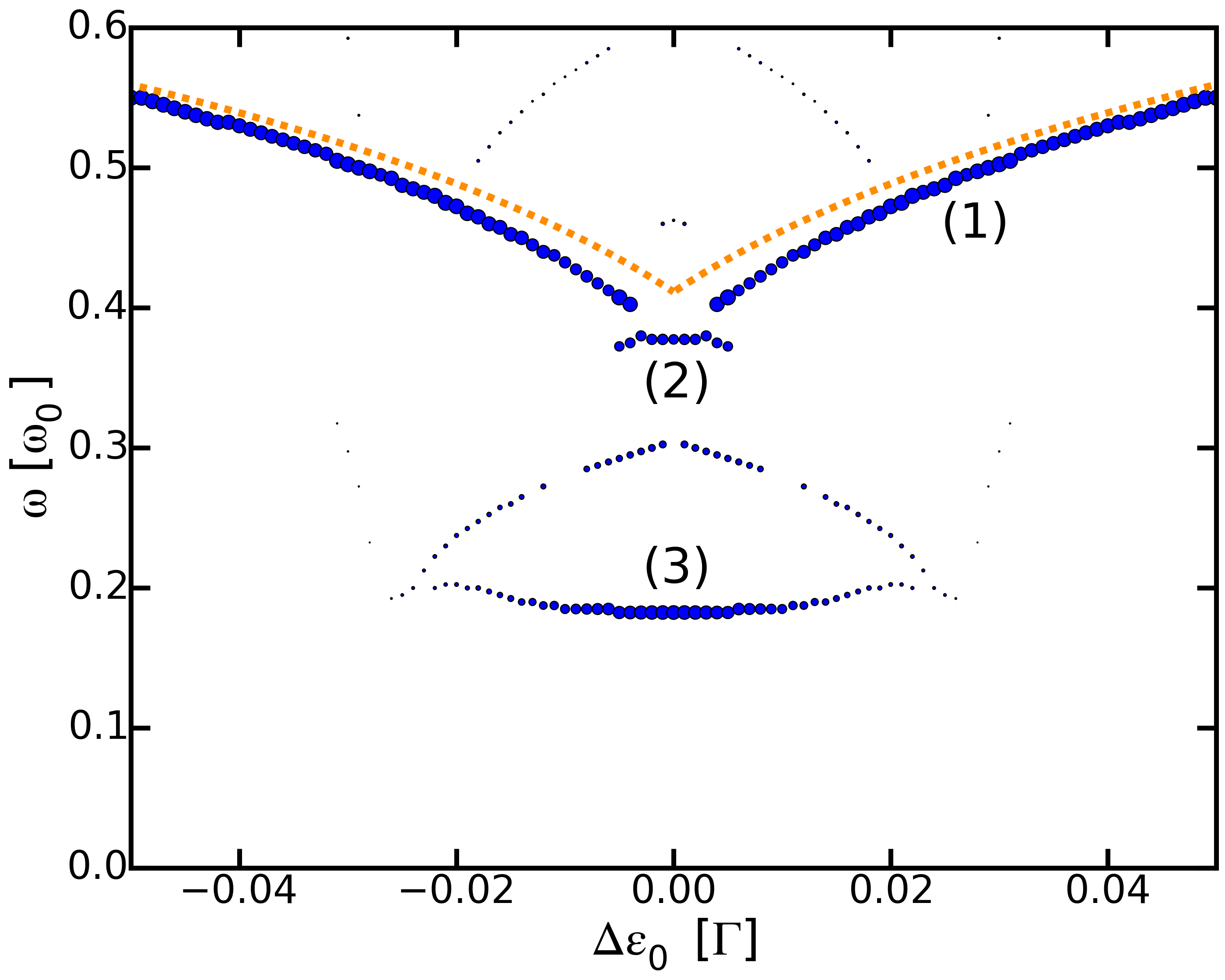}
 \caption{(Color online)
 Same as Fig. \ref{figVg1} but for $\EP = 1.1 \EC$, corresponding to the bistable region.
 The orange dotted lines indicate the numerical solution of the mean field \refe{y0eq} and
 \refe{omegamtau}.
 The dots indicate the positions of the dominant maxima in the $S_{xx}(\omega)$ spectrum, with a dot size proportional
to the peak height.
 }
 \label{figVg2}
\end{figure}

The results for $\EP<\EC$ and $\EP>\EC$ are shown in Figs. \ref{figVg1} and \ref{figVg2}. 
The behavior discussed by simple analytical argument agrees with what is found by solving
numerically \refe{y0eq}  and substituting the result into \refe{omegamtau} (dotted lines in both figures). 
As explained before the calculation of the resonating frequency by this method gives only an indication of 
the mechanical dynamics, that is actually much more complex. 
In the next sub-section we thus discuss the form of the displacement and current spectrum taking
into account the effect of the fluctuations.

\subsection{The displacement- and current-spectra as a function of $\Edt$}
\label{Sec_mode_softening_fluctuations}

Using the numerical methods described above we obtain the displacement and current spectrum.
For $\EP\leq \EC$ the displacement spectrum shows a clear peak at frequencies that are 
very close to the results obtained by the analytical method. 
The position of the peak is shown by the continuous lines in \reff{figVg1}. 
For weak coupling the agreement is essentially perfect. 
For $\EP=\EC$ one see instead that the peak remains at a higher energy. 
As we found for the case $\Edt=0$ in Ref.~\onlinecite{micchi_mechanical_2015} the strong 
non-linear form ($x^4$) of the potential at that point is at the origin of this difference.
There the spectrum has a form:
\begin{equation}
 	S_{xx}(\omega) = S_{xx}(\omega_M) \left(\frac{\omega}{\omega_M}\right)^4 e^{-\left[\left(\frac{\omega}{\omega_M}\right)^4 - 1\right]}\, ,
 \label{eq:SxxCriticality}
\end{equation}
with $\omega_M\approx 1.2 \omega_0  (T_\text{eff}/\Gamma)^{1/4}$.
The width of the peak is due to the phase fluctuation, and it is so wide since $\omega(E)$ 
given by \refe{omegaE} vanishes for $E=0$.

\begin{figure*}
 \includegraphics[width = \textwidth]{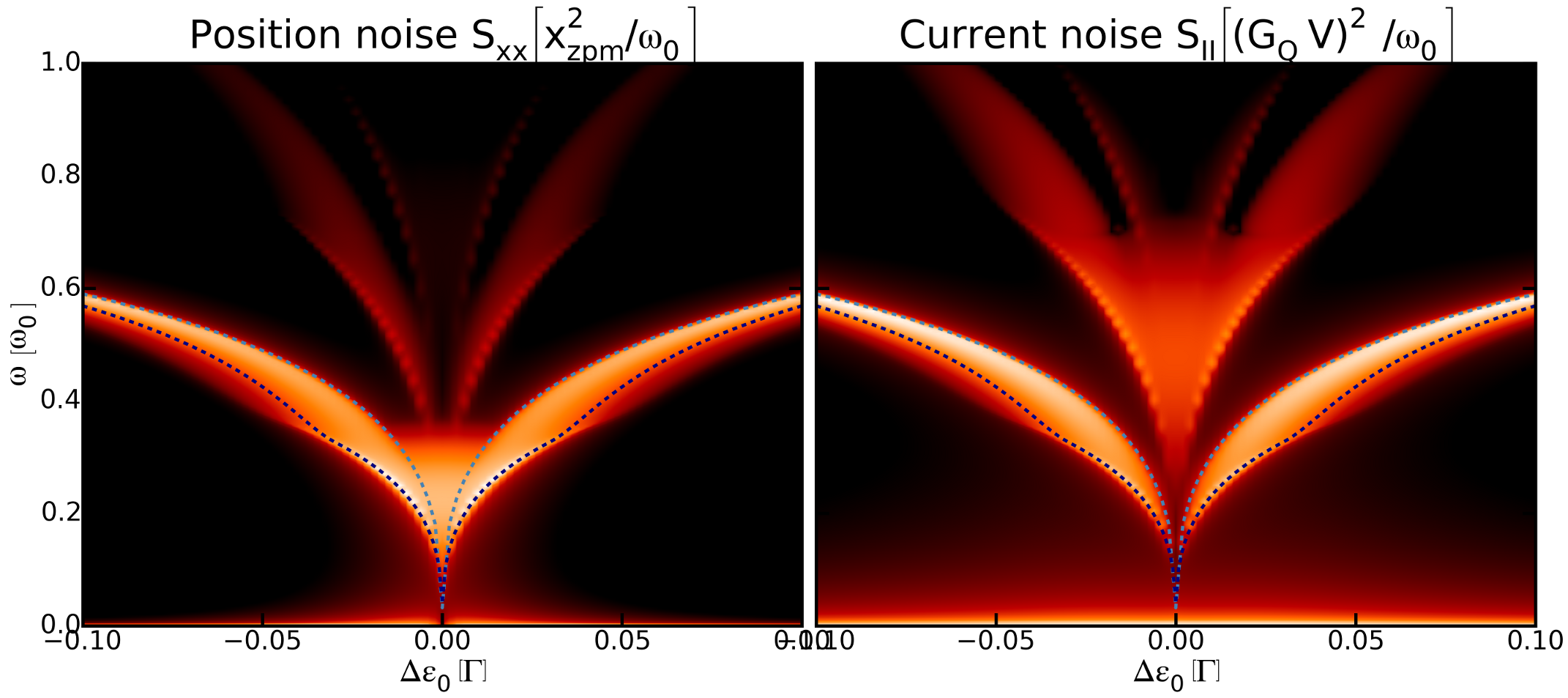}
 \caption{(Color online) Density plot of the displacement-spectrum $S_{xx}$ (left panel) 
and current spectrum $S_{II}$ (right panel) as a function of dot-level detuning 
$\Delta\Ed$ and frequency $\omega$.
The dashed light-blue line shows the position of $\omega_m$ from the solution of \refe{y0eq} and \refe{omegamtau}.
The dashed dark-blue line shows the position of $\omega(E_0)$ at which the dispersion relation has a minimum.
The maximum value in the density scale has been chosen according to the most important finite-frequency peak, and the minimum value has been chosen 4 orders of magnitude below. The plot is presented in logarithmic scale.
Note that a zero-frequency noise peak is present and is typically several orders of magnitude 
above the max of the scale, especially in the current plot.
The parameters are the same of the previous two figures.}
 \label{fig:gate_spectra}
\end{figure*}

At finite value of $\Edt$, the effective potential of the oscillator gets an asymmetric minimum (see lower panel of \reff{figVg1})
which is responsible for a non-vanishing $\omega(E)$ at $E=0$, as given by \refe{omegamtau}.
This reduces the range over which the frequency can vary.
As a consequence, at sufficiently large detuning $\Edt \ge 0.1 \Gamma$, the peak quite rapidly follows again the mean-field prediction
shown by the light-blue dashed curve in \reff{fig:gate_spectra} (left panel).
For smaller dot-level detuning $\Edt < 0.05 \Gamma$, the spectrum becomes more complex, with the emergence
of a secondary peak that is not described by the mean-field mode softening.
This is due to the existence of a precise value of the energy $E_0 > 0$ (dependent on $\Edt$)
for which the dispersion relation $\omega(E)$ of the oscillator develops a minimum.
For $\omega \approx \omega(E_0)$ there is a square-root divergence of the
spectrum $S_{xx}(\omega) \propto E_0 \sqrt{\omega''(E_0)/2\left\lbrack \omega - \omega(E_0) \right\rbrack}$.
This divergence is further widened by dissipation and leads to a narrow peak,
as shown by the dark-blue dashed curve in \reff{fig:gate_spectra} (left panel).
The double-peak feature rapidly disappears close to criticality,
where one recovers the spectrum described by \refe{eq:SxxCriticality}.
%

%
%
%
%
%

Particularly interesting is the case $\EP>\EC$.
As we have seen analytically in \refe{eq:cusp}, we expect a cusp in the mean-field peak position close to $\Edt=0$.
This is also found numerically, as it can be seen from \reff{figVg2}, branch (1), at least for $\Edt$ not too small.
When we look more in details close to $\Edt=0$, we find two minima in the effective potential, one stable and one meta-stable.
We assign the peaks in branch (1) to oscillations around the stable minimum.
The contribution of the meta-stable state rapidly vanishes due to its population, exponentially small in the regime $T_\text{eff} \ll \Gamma$.
It can still be seen close to its merging with the stable-minimum peak, arising at branch (2).
However, its presence is fundamental to generate branch (3), that is associated to oscillations around both minima and of which frequency is roughly half the frequency of branch (1).

We can calculate also the current spectrum. We show in \reff{fig:gate_spectra} the 
result for $\EP=\EC$. 
In order to understand the relation with the displacement spectrum one can generalize
\refe{eq:SII_2}:
\begin{equation}
 \begin{aligned}
  {S_{II}(t)\over (F_0/\Gamma)^2}  
  \approx& \left( G_Q V \right)^2 \tau^3 \Big\{  4(1-\tau)S_{xx}(t) \\
  & + 
  \frac{\Gamma^2}{F_0^2} \tau(3-4\tau)^2
  \left[ \left\langle y^2(t)y^2(0)\right\rangle - \left\langle y^4(0)\right\rangle \right] \\
  & +
  \frac{\Gamma^2}{F_0^2} 16(1-\tau)(1-2\tau)
  \left[ \left\langle y(t)y^3(0)\right\rangle - \left\langle y^4(0)\right\rangle \right]  \Big\}
 \end{aligned}
 \label{eq:SII_asym}
\end{equation}

The first term in ~\refe{eq:SII_asym} is proportional to the mechanical noise $S_{xx}(t)$.
It dominates the spectral signal for detuned dot-level position ($\Delta \Ed \ne 0$ and $\tau \ne 1$) and
is at the origin of the double V-shape seen in Fig.~\ref{fig:gate_spectra}-(b).
The second and third terms in ~\refe{eq:SII_asym} are related to the fourth-order correlation functions of the oscillator position.
The second term dominates the spectrum in the region close to the critical point ($\Delta \Ed =0$ and $\tau=1$).
The line-shape in this regime is well reproduced by ~\refe{eq:SII_regular} giving the regular part of the spectral density.
Finally, we remark the presence of a low-frequency noise due to the singular part of the spectral density written in \refe{eq:SII_singular}.

To conclude, we have seen that the transition to the bistable phase can be detected by studying the displacement
or current spectrum as a function of the detuning of the single-electron level.
Specifically the shape of the softening depends strongly on the value of the interaction.
A cusp should appear for coupling larger than the critical value. 
Clearly, fluctuations smoothen the transition, but the characteristic dependence on the bias voltage
of the structures should be also a valuable indication of the true nature of the transition. 


\section{The transition at finite temperature}
\label{sec:temperature}

In this section we discuss the behavior of the system when the electronic leads are
kept at finite temperature.
Two different manifestations of the finite temperature are:
({\em i}) The distribution of the electronic population of the metallic leads modifies the 
transport properties of the quantum dot and in particular its average electronic occupation 
[$\langle n(x)\rangle$]
that is responsible for the effective force acting on the oscillator.
({\em ii}) Thermal electronic fluctuations induce equilibrium fluctuations of $n(x)$ thus leading to a
stochastic force that heats the oscillator even for $V=0$.

\subsection{Temperature dependence of the effective potential}
\label{Sec_mode_softening_Temperature}
\begin{figure}
\includegraphics[width=1.0\linewidth]{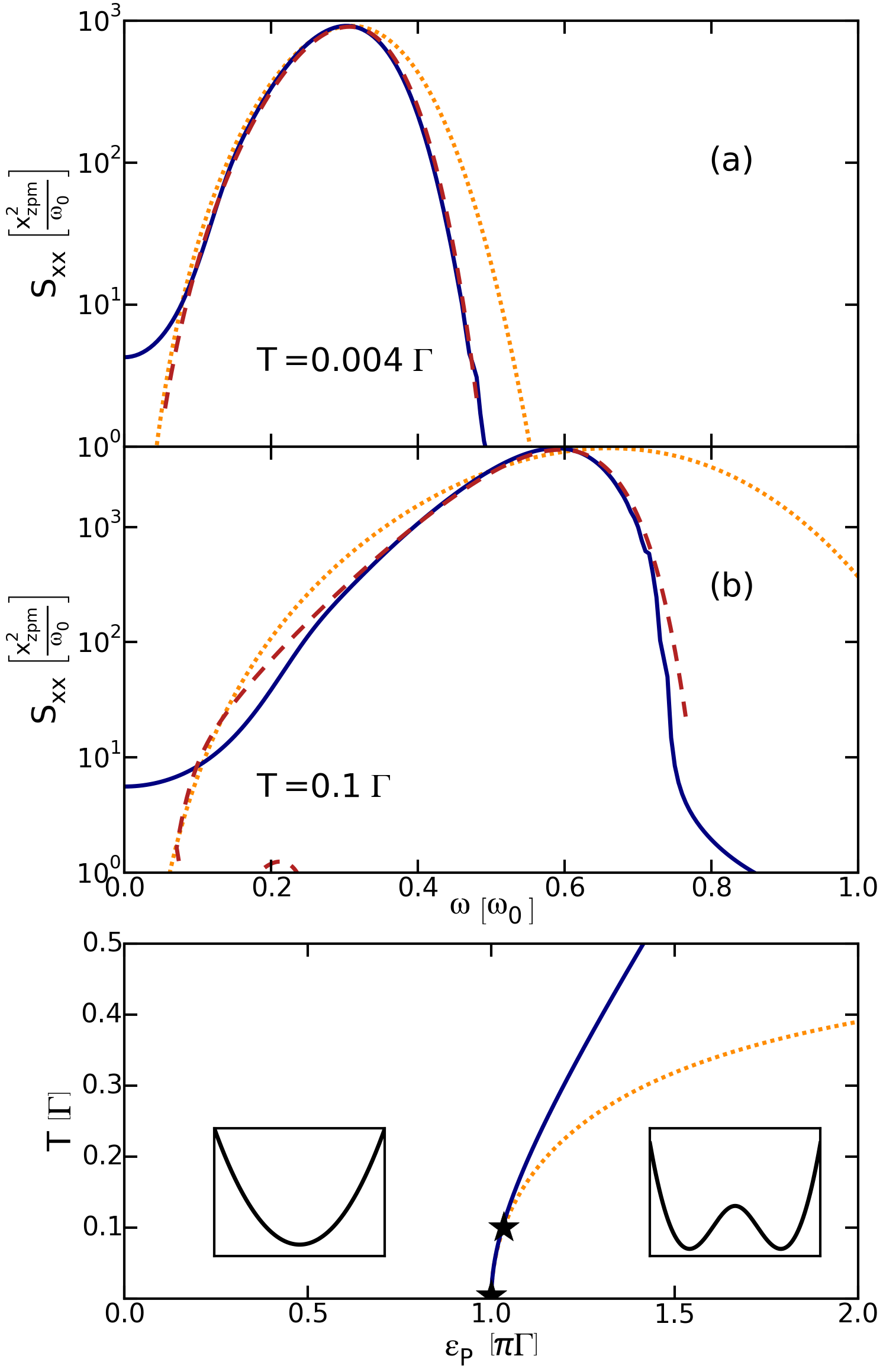}
\caption{(Color online)
Top panels:  
displacement spectrum $S_{xx}(\omega)$ for (a)
$T/\Gamma=4\cdot 10^{-3}$ and (b) $T/\Gamma=0.1$.
Solid (blue) lines indicate the numerical results from \refe{eq:Sii_pistolesi}, 
while dotted (orange) lines stand for the analytical low-temperature approximation 
of \refe{eq:SxxCriticality} with renormalized parameters.
The dashed (red) lines correspond to evaluating numerically \refe{eq:SII_3} 
for the mechanical noise, using the exact expression for the effective potential.
Bottom panel:
Phase diagram of the system in the plane $T$-$\EP$. 
The bistability is present only in the area to the right of the transition, as is 
illustrated by the potential energy (see the insets). 
The analytical transition line for $T \ll \Gamma$ is shown in dotted (orange) line following \refe{criticalLowT}.
The two stars indicate the values at which the top panels are calculated.
The parameters are the same of previous figures.
}
\label{fig:spectrum_t}
\end{figure}

\refe{eq:FAD_electronicN} gives the expression of the average electronic occupation of the dot. 
At finite temperature it can be expressed in terms of the digamma function ($\Psi$):
\beq
	n={1\over 2}+ {1\over 2\pi } {\rm Im} \sum_{\alpha=L,R} 
	\Psi \left( {1\over 2}+{\Gamma + i (\mu_\alpha-\epsilon_0+F_0 x) \over 2 \pi T} \right)
	\,.
\eeq
We are interested in studying how the transition is modified by the temperature. 
For this we begin by considering the behavior of $\omega_m^2$
as defined by \refe{defomegam}.
The change of sign of $\omega_m^2$ indicates that the stationary point is no more stable.
For simplicity we focus on the $V=0$ case. 
The critical value of the coupling constant for which $\omega_m^2=0$ reads then:
\beq
	\EC(T)=-\left( { \partial n \over \partial \epsilon_0 }\right)^{-1}={2 \pi^2 T\over \Psi^{(1)}(1/2+\Gamma/2\pi T)} ,
		\label{criticalT}
\eeq
with $\Psi^{(n)}(z)=d^n \Psi(z)/dz^n$ the $n$-derivative of the digamma function 
and we consider only the case 
$\epsilon_0=\EP/2$. 
Using known properties of the digamma function
[$\Psi^{(1)}(1/2)=\pi^2/2$, $\Psi(z)\approx 1/z+1/(2z^2)+1/(6z^3)+\dots$ for large $z$],\cite{abramowitz_handbook_2013}
one finds the asymptotic  $T \gg \Gamma$ behavior
\beq
	\EC(T \gg \Gamma) = {4T \over 1+\Psi^{(2)}(1/2)\Gamma/(T \pi^3)}
\eeq
and the low temperature one 
\beq
	\EC(T \ll \Gamma) = \pi \Gamma\left( 1+ {\pi^2\over 3}{T^2 \over \Gamma^2}\right)
		\label{criticalLowT}
\eeq
(the last expression can also be obtained by Sommerfeld expansion).
One thus finds that the transition can be in principle observed also at high temperatures, 
provided a coupling constant of the order of $4T$. 
By a comparison with the result (\ref{criticalV}) for the bias voltage dependence of $\EC$, one sees that 
the temperature (for $T\gg \Gamma$) is less effective in the renormalization of the spring constant. 
Clearly the same expressions allow one to study the evolution of $\omega_m$:
\beq
	{\omega_m^2 \over \omega_0^2}
	=
	1-{\EP\over \EC(T)}
	.
\eeq
At mean field level the temperature simply changes the critical value of the coupling.
The critical line (\ref{criticalT}) is shown in Fig.~\ref{fig:spectrum_t} lower panel full line. 
The dotted (orange) curve shows the low temperature approximation (\ref{criticalLowT}).

To conclude the discussion we remark that not only the coefficient of the quadratic term in the 
effective potential is modified by the temperature, but also the higher order ones.
Using the fact that $\Psi(z)=d \log \mathbf{\Gamma}(z)/dz$ we can write an analytical expression also 
for the effective potential:
\beq
	U(y)={\Gamma^2 \over 2 \EP} y^2+2 T {\rm Re} \log  
	\mathbf{\Gamma}\left( {1 \over 2} + {\Gamma \over 2 \pi T}(1+iy)\right) 
	\label{UfiniteT}
	,
\eeq
were we have used the dimensionless position $y$ introduced above.
One can verify using the asymptotic behavior for large $z$ of
$\log \mathbf{\Gamma}= z \log z-z-\log(z/2\pi)/2+\dots$,\cite{abramowitz_handbook_2013} 
that \refe{UfiniteT} reduces to \refe{UzeroT} for $T\ll \Gamma$. 
The generic $n$-term of the expansion of $U(y)=\sum_{n=2}^\infty U_n y^n $ reads: 
\beq
	U_n 
	=
	\delta_{n,2} {\Gamma^2 \over 2 \EP}
	+
	{\Gamma \over \pi n!}  \Psi^{(n-1)}\left( {1\over 2}+{\Gamma\over 2\pi T}  \right)
	\left({\Gamma\over 2 \pi T}\right)^{n-1}
	\label{UnExact}
	,
\eeq
with $U_n$ non vanishing only for $n$ even.  
This gives immediately that the quartic term $U_4$ is always positive and 
vanishes for large $T$ as $1/T^4$.
In general the effective potential induced by the electrons vanishes for large temperature,
since in this limit the average occupation of the dot, $n$, depends very weakly on the position of the level.
In the opposite limit, for $T\ll \Gamma$, one finds for the quartic term of the expansion of $U(y)$ 
$U_4=(\Gamma/12\pi)\left[1-2 \pi^2(T/\Gamma)^2\right] y^4$.
%

%
%
%
%
%
%
%

\subsection{Effect of thermal fluctuations on mechanical noise}

\label{Sec_mode_softening_fluctuations_temperature}

In thermal equilibrium ($V=0$) the distribution function is always given by \refe{GibbsP} with 
$T_{\rm eff}=T$. 
One can thus obtain the spectrum of both $I$ and $x$ from expressions like \refe{eq:SII_3}.
The only difference with the previous case is the form of the effective potential $U$ that leads to a different 
expression for $\omega(E)$ and $x_E(t)$. 

Specifically at the critical line the quadratic term vanishes and as far as the quartic term dominates 
over the higher order terms the universal line-shape for $S_{xx}(\omega)$ given in  \refe{eq:SxxCriticality} holds.
The only difference is the value of the maximum of the peak $\omega_M$
\begin{equation}
 \omega_M(T) \approx 1.212 \omega_0\, \left(\frac{T}{\Gamma}\right)^\frac{1}{4}
 \left[ 1 - {\pi^2\over 2} \left( \frac{T}{\Gamma} \right)^2 \right]
\end{equation}
Hence with increasing temperature, the maximum of the spectral line $S_{xx}(\omega)$ moves toward higher frequencies while its width increases.

In Fig.~\ref{fig:spectrum_t} upper panels we plot the displacement spectrum $S_{xx}(\omega)$ obtained at two different temperatures $T/\Gamma=4 \cdot 10^{-4}$ [panel (a)] and $T/\Gamma=0.1$ [panel (b)]:
The (orange) dotted line presents the analytical results given by \refe{eq:SxxCriticality} and the (blue) solid line presents the full numeric calculations given by \refe{eq:Sii_pistolesi}.
We observe that the analytical results are qualitatively consistent with the numerics: 
Namely, we find a shift of the peak toward higher frequencies and an enlarged broadening of the 
resonance upon increasing temperature.
However a quantitative discrepancy is evident, especially at higher temperature [panel (b)].
This is due to the fact that by increasing the temperature the fluctuations increase and at 
some point the sixth and higher order terms cannot be neglected anymore. 
Let's assume that the distribution function is $P\sim e^{-U_4 x^4/T}$.
We can then evaluate the contribution of any term in the expansion of $U(y)$:
\beq
	U_n \langle x^n \rangle = \mathbf{\Gamma}(1/4+n/4) \left({T\over U_4 }\right)^{n/4}
		\,.
\eeq
We thus find that $U_n \langle x^n \rangle/(U_4 \langle x^4 \rangle) \sim (U_n/U_4)(T/U_4)^{(n-4)/4}$.
From expression (\ref{UnExact}) one finds that for $T\rightarrow 0$ the coefficients $U_n \sim \Gamma$.
This gives the condition $(T/\Gamma)^{(n-4)/4} \ll 1$ in order to neglect the term of order $n$ with respect to the 
term of order 4. 
For the case in panel (b) the condition for the sixth terms reads $\sim \sqrt{T/\Gamma} \approx 0.3$. 
We verified that evaluating numerically the expression equivalent to \refe{eq:SII_3} for $S_{xx}$ 
taking into account the exact form of the potential (see dashed lines in the figure) 
we recover the results of the full numerical solution.

\section{The effects of a dissipative coupling}
\label{sec:dissipation}

Up to this moment, we considered the coupling with the electrons to be the only source of dissipation of the system.
In this section we consider the effect of a coupling to a bath at the same temperature of the electronic degrees of 
freedom, but with an independent coupling constant modeled by the damping rate $\gamma_\text{e}$.
The fluctuations induced by this coupling satisfy the fluctuation-dissipation theorem, leading to a 
modification of the Fokker-Planck \refe{eq:FP} as follows:
\beq
	{A\over m} \rightarrow {A\over m}+\gamma_\text{e}, \qquad 
	D \rightarrow D+2 \gamma_\text{e} mT \, .
\eeq
In the limit $T,V \ll \Gamma$ the stationary solution is still of the Gibbs form (\ref{GibbsP}) with effective 
temperature:
\beq
 	T_\text{eff} (\gamma_\text{e}) =  {D+2m \gamma_\text{e} T \over 2 A + 2 m\gamma_\text{e}}
 		= {T_\text{eff} (0)+T m\gamma_\text{e}/A \over 1+m \gamma_\text{e}/A}
 		\,.
\eeq
From this expression one can conclude that for $m \gamma_\text{e}/A\ll 1$ and 
$T m\gamma_\text{e}/A \ll T_\text{eff}(0)$ the effect of an additional dissipative channel 
can be neglected. 
Since $T_\text{eff}(0)\geq T$ as given by \refe{DefTeff}, the first condition is sufficient to 
neglect the effect of fluctuations. 
%

%
%
%
%
%
%
%
\begin{figure}
 \includegraphics[width=1.0\linewidth]{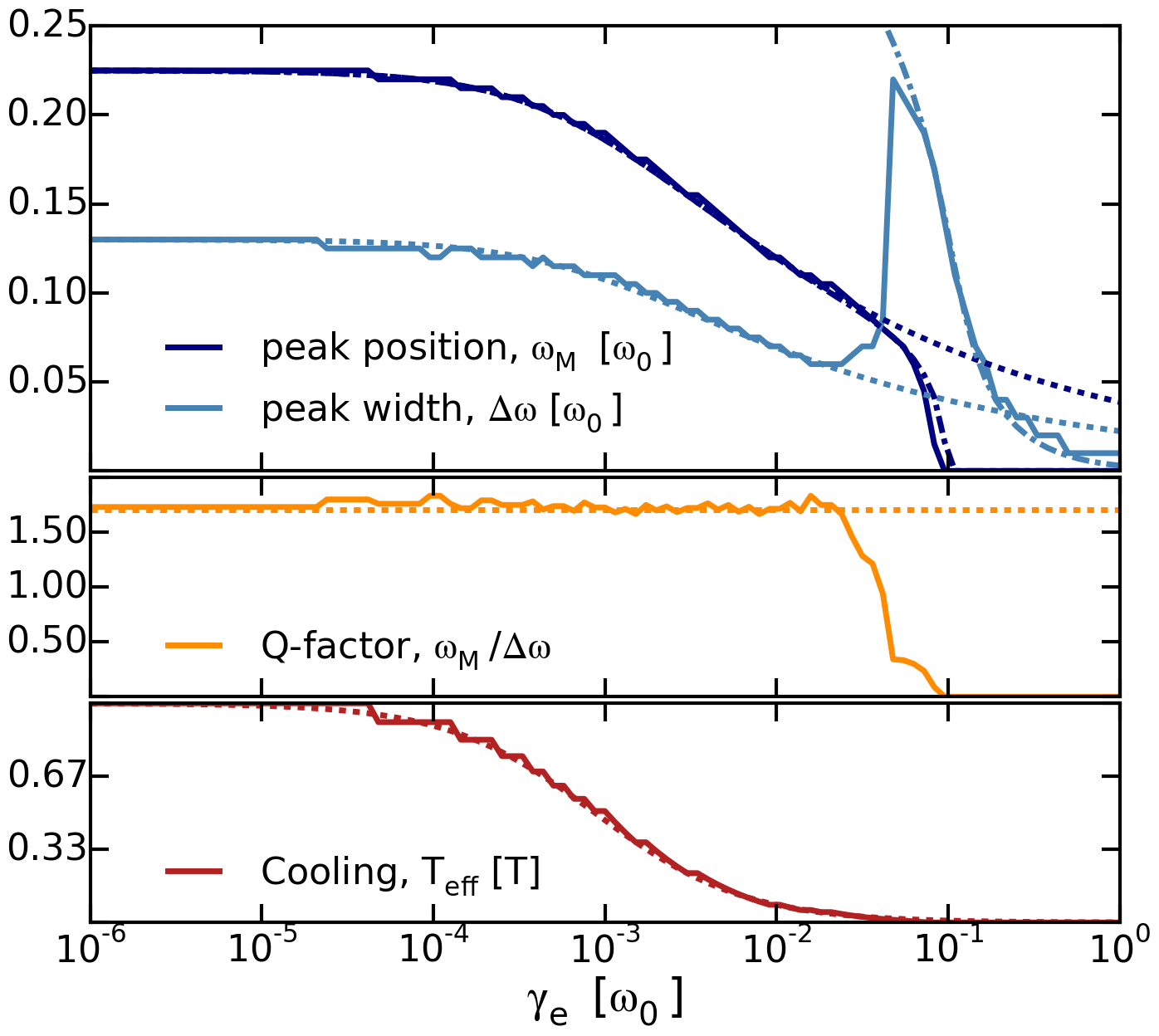}
 \caption{
(Color online)
We plot as a function of the damping constant $\gamma_\text{e}$ the position and the width of the maximum of 
$S_{xx}(\omega)$ (upper panel) , the quality factor (middle panel), and the effective temperature $T_\text{eff}$ 
(lower panel). 
Solid lines represent the numerical results obtained by solving the Fokker-Planck equation.
The dashed lines in the upper and central panel indicate the evolution of $\omega_m$ and $\Delta \omega$ 
given by the analytical expression $\omega_M=1.2 \omega_0 (T_\text{eff}/\Gamma)^{1/4}$ and
$Q=\omega_M/\Delta \omega=1.71$ valid at criticality.
The dot-dashed line in the upper panel is given by \refe{omega2def} and \refe{omega1}.
The parameters are $\Gamma = 1000 \, \omega_0$, $\EP = \EC$, $T=0$, and $V/\Gamma = 5 \cdot 10^{-3}$.
}
\label{fig:ext_dissipation}
\end{figure}

When $m \gamma_\text{e}/A$ is not small there are two possibilities. 
If $T\gg V$, then one simply finds $T_\text{eff} (\gamma_\text{e}) =T_\text{eff} (0)=T$, independently 
of  $\gamma_\text{e}$. 
More interesting is the opposite limit, of $T \ll V$, in this case $T_\text{eff} (0)=V/4$ and we find:
\beq
 	T_\text{eff} (\gamma_\text{e}) =   {T_\text{eff} (0) \over   1+m \gamma_\text{e}/A} .
 	\label{RenTeff}
\eeq
The effect of the dissipation is then to cool down the oscillator. 
This is not surprising: the presence of a voltage bias induces heating and by coupling
the mechanical oscillator to the cold environment one can reduce the effective temperature.
Thus we find that for the observation of the transition the presence of an additional 
(and in general unavoidable) dissipation, is either uninfluential or an advantage (in the case $T\ll V$). 

In order to see this we solve numerically the problem and show 
in Fig.~\ref{fig:ext_dissipation} as a function of $\gamma_\text{e}$ for $T=0$
the dependence of the main peak position of $S_{xx}(\omega)$, its width, its quality factor $Q$, and the 
effective temperature of the system  $T_\text{eff} $.
One clearly sees that increasing the dissipation reduces the position of the main peak and its width,
but the ratio remains perfectly constant, as shown by the evolution of the quality factor. 
All the dependence can be explained by the renormalization of the effective temperature 
predicted by \refe{RenTeff}.

The only point that requires some additional explanation is the way the peak becomes 
overdamped when $\gamma_\text{e}\approx \omega_M$.
We found that this can be understood in terms of a simple model. 
Let's consider the displacement spectrum for a damped harmonic oscillator 
of frequency $\omega_\text{t}$ and damping $\gamma$ driven by a force noise $\xi(t)$:
$S_{xx}(\omega)=(|\xi(\omega)|^2/m^2) [(\omega^2-\omega_\text{t}^2)^2+\omega^2 \gamma^2]^{-1}$.
Assuming a white noise $\xi(\omega)\approx \xi(0)$ we find that $S_{xx}(\omega)$ has
two maxima at $\omega =\pm \omega_1$ with 
\beq
	\omega_	1=(\omega_\text{t}^2-\gamma^2/2)^{1/2}
	\label{omega1}
\eeq
for $\gamma<\sqrt 2 \omega_\text{t}$ and a single maximum at $\omega=0$ for $\gamma \geq \sqrt 2 \omega_\text{t}$.
Assuming that $\omega_M$ plays the role of $\omega_\text{t}$ we plot as a 
dot-dashed line in the upper panel of Fig.~\ref{fig:ext_dissipation} 
the expected behavior of the maximum on $S_{xx}(\omega)$ as predicted by \refe{omega1} .
This agrees remarkably well with the full numerics, even if the origin of the 
peak is due to quartic fluctuations. 
We think that the reason is that the dissipation dominates, thus it is not very important the origin of the peak. 
Pushing even further the model in the overdamped regime we looked at the evolution of the peak by calculating the value of 
$\omega$ for which $S_{xx}(\omega_2) = S_{xx}(0)/2$.
We find
\beq
	\omega_2 = 
	\left[ 
	\omega_\text{t}^2 - \frac{\gamma^2}{2} + 
	\sqrt{\left(\omega_\text{t}^2 - \frac{\gamma^2}{2}\right)^2 + \omega_\text{t}^4}
	\right]^{1/2}
	\label{omega2def}
\eeq
We compare then $2\omega_2$ with the full width half maximum found numerically. 
Note that numerically when the peak has only one side that allows to find the half-value we 
simply double the distance from the maximum to this frequency. 
Again a comparison of the linear model and the full numerics works remarkably well. 

\section{Conclusions}
\label{sec:conclusions}

In this paper we have studied the transition to a mechanical bistability induced by the strong coupling between 
a mechanical degree of freedom and the charge in a quantum dot.
We have considered the experimentally relevant regime of $\Gamma\gg \omega_0$.
We have studied the phase diagram of the problem as a function of the bias voltage (cf. \reff{fig:phase_diagram}), 
the temperature (cf. \reff{fig:spectrum_t}),
and the gate voltage (cf.~Figs.~\ref{figVg1} and \ref{figVg2}). 
The critical value of the coupling $\EC$ to observe the transition depends on the voltage quadratically (cf. \refe{criticalV}) 
and on the temperature linearly (cf. \refe{criticalT}). 
Since reaching large coupling is a difficult experimental problem, the ideal situation for the observation of the transition 
is the low temperature and low bias voltage case for which $\EC=\pi \Gamma$.
We investigated this limit in a previous publication\cite{micchi_mechanical_2015}. 
One of the main message was that the mechanical degrees of freedom had a much stronger response at the transition 
than the electronic ones. 
To clarify better this point in the present paper we considered in details the behavior of the conductance (cf.~\reff{fig:conductance}).
We find that it has a discontinuous behavior at exactly vanishing temperature and bias voltage, but in practice
it becomes smooth very rapidly when $V>0$ and it has a slow power low decrease for large coupling.
Thus for sufficiently large coupling this leads to a well defined blockade,\cite{pistolesi_self-consistent_2008}
but in the range relevant for the present experiments the conductance does not provide an unambiguous 
proof of the bistability. 
For this reason we investigate in details the quantities that are directly related to the mechanical degrees of freedom. 
The main effect at the transition is the softening of the mechanical mode as predicted by the behavior of the
electronic effective potential.
This reflects in the response function to an external driving force, in the ring-down time, and in particular in the 
displacement fluctuation spectrum $S_{xx}(\omega)$.\cite{micchi_mechanical_2015} 
Here we have investigated in details the behavior of the current-spectral density $S_{II}(\omega)$.
This quantity has been measured for suspended carbon nanotubes.\cite{moser_nanotube_2014}
We found that measurement of $S_{II}(\omega)$ can give a direct access to $S_{xx}(\omega)$, but in a simple way 
only deep in the bistable phase. 
Actually in the most interesting case of resonant tunnelling ($\Delta \epsilon_0=0$) the 
transparency of the junction depends quadratically on the displacement, leading to a relation \ref{eq:SIIPerturbative_3} 
between $S_{II}$ and $S_{xx}$, that breaks down at the transition due to the strong non-linearity. 
One can nevertheless obtain the behavior of the current noise, that has similar remarkable features of
$S_{xx}$ at the transition [cf. \refe{eq:SII_regular} and (\ref{eq:SII_singular})].
The behavior of the peak position ($\omega_M$) and of its width constitutes robust fingerprints of the transition.

We investigated the role of a detuning of the dot level $\Delta \Ed$ and of a finite temperature $T$ on this critical behavior.
We showed that the dependence of the mode softening with $\Delta \Ed$ is extremely sharp close Fto criticality,
scaling with $\left(3|\Delta \Ed|/\Gamma\right)^{1/3}$.
This in striking contrast with the Lorentzian behavior expected in the weak-coupling limit and actually observed in all
current experiments.
We found that the main effect of the temperature is to change the value of the critical coupling needed
 to observe the transition and, of course, to increase the fluctuations by changing the effective temperature. 
Finally we showed that the effect of a dissipation not due to the electrons is actually in general useful 
to keep the oscillator cold in the case the voltage bias is larger than the electronic temperature. 

We gave thus a global picture of the transition, investigating the main physical quantities, and showing that several features can 
be used to characterize without ambiguity the transition to the bistable state.
These results give clear indications  opening the way to the observation of this phenomenon with state-of-the-art experiments.

The experimental observation of the transition would indicate that one has entered a completely new regime, where the 
interaction due to a single electron has a dramatic effect on transport and mechanical behavior.
In molecular devices this kind of effects might have been observed,\cite{tao2006electron, lortscher2006reversible, PhysRevLett.96.156106, PhysRevLett.85.1918}
but in these systems is nearly impossible to tune the electron-phonon coupling.
Observation of the transition in nano-electro mechanical systems, like the suspended carbon nanotubes, would then open the way to controlled investigations of the strong coupling regime. 
The increase in the coupling leads naturally to an increase in the sensitivity of the detection, with improvement in the 
sensitivity of the device (for instance for mass or force detection).
From the fundamental point of view, a better understanding of the behavior of mechanical resonators in the ultra-strong coupling regime could pave the way for testing decoherence \cite{leggett_testing_2002} as well as reaching the ultimate
limits in the control of mechanical motion at the nano-scale. \cite{poot_mechanical_2012}

From the theoretical point of view there are still open questions. 
At the moment it is difficult to perform experiment in the quantum regime 
($\omega_0 \sim T$), but the observation of very high frequency mechanical resonators,\cite{laird_high_2011}
leaves open the possibility of reaching at some point this limit and thus should motivate theoretical studies.
An even more stringent question is opened by a recent publication,\cite{klatt_kondo_2015}
where by mapping the quantum problem to an effective Kondo problem it was shown that the bistability
in the quantum regime may be washed out.
The study was performed in equilibrium and focussed only on the probability distribution. 
An investigation of the response functions $S_{xx}$ in that regime would be extremely interesting 
in order to clarify the expected evolution of measurable quantities at the transition for low temperatures.

\section*{Acknowledgements}

The authors thank S. Ilani and V. Golovach for useful discussions.

\appendix


\section{Derivation of $\langle n \rangle$ and $\langle n(t)n(0)\rangle$} 

\label{app:langevin}

In this appendix we present a derivation of the 
$\langle n \rangle$ and $\langle n(t)n(0)\rangle$
based on the method known as input-output theory 
widely used in quantum optics (see for instance Ref.~\onlinecite{walls_quantum_2008})
for bosonic fields, but much less used to describe fermions.
We consider the Hamiltonian (\ref{eq:Hamiltonian})
for given $\epsilon_d=\epsilon_0-F_0 x$,
$x$ here is a constant $c$-number. 
The Hamiltonian is quadratic in the fermionic operators.
One can then calculate exactly all correlation functions. 
We begin by writing the equation of motions for the fermionic 
operators in the Heisenberg representation:
\begin{equation}
        \begin{cases}
                \dot d(t) = -i \epsilon_d \ d(t) - i \sum_{\alpha k} t_\alpha^* \ c_{\alpha k}(t) \\
                \dot c_{\alpha k}(t) = -i \epsilon_{\alpha k} \ c_{\alpha k}(t) - i t_\alpha \ d(t) 
        \end{cases}
        \label{eq:motion_io}
        \, .
\end{equation}
It is convenient to introduce new operators:
$\tilde d(t) = \exp(i\epsilon_d t) d(t)$ and $\tilde c_{\alpha k}(t) = \exp(i\epsilon_d t) c_{\alpha k}(t)$.
In terms of these operators we can solve the second of the two equations
(\ref{eq:motion_io}):
\begin{equation}
\label{eq:ctilde}
        \tilde c_{\alpha k} (t) = -i t_\alpha \int_0^t{e^{-i\Delta_{\alpha k}(t-t')} \tilde d (t') dt'} 
        + e^{-i\Delta_{\alpha k}t} \tilde c_{\alpha k} (0) \, ,
\end{equation}
where we have defined the energy difference 
$\Delta_{\alpha k} = \epsilon_{\alpha k} - \epsilon_d$ and $\tilde c_{\alpha k} (0)$ is 
the initial condition for the equation. 
Substituting the form (\ref{eq:ctilde}) into \refe{eq:motion_io} we have:
\begin{equation}\label{eq:dtilde_diffeq}
        \dot{\tilde{d}}(t) = -\sum_{\alpha k} \left[|t_\alpha|^2 
        \!\! \int_0^t  \!\!\!  {e^{-i\Delta_{\alpha k}(t-t')} 
        \tilde d (t') dt'}  
        +i t_\alpha^*  e^{-i\Delta_{\alpha k}t} \tilde c_{\alpha k} (0)\right]
        \,.
\end{equation}
Since the leads are metallic the summation over $k$ can be replaced by an integration over the energy 
by defining the density of states of the system:
 $\rho_\alpha (\nu_\alpha) = \sum_k \delta\left(\nu_\alpha - \Delta_{\alpha k}\right)$.
 In the wide-band approximation we neglect the energy dependence of the density of states and we obtain:
\begin{equation*}
        \sum_k |t_\alpha|^2 \int_0^t dt' e^{-i\Delta_{\alpha k}(t-t')} = 2\Gamma_\alpha \ \delta(t - t').
\end{equation*}
where we have introduced $\Gamma_\alpha$ as defined in section~\ref{sec:model}.
It is convenient to define an incoming field:
\beq
 \label{eq:incident_field}
        \tilde c_{\alpha, \text{in}}(t) = \sum_k e^{-i \Delta_{\alpha k} t} c_{\alpha k} (0)
        .
\eeq
Defining  $t_\alpha = |t_\alpha| \ \exp(-i\phi_\alpha)$ we have:
\begin{equation}
	\label{eq:dtilde_wba}
        \tilde{\dot d}(t) = -\Gamma \tilde d (t) 
        - \sum_\alpha i\ e^{-i\phi_\alpha} ({\Gamma_\alpha}/{\pi\ \rho_\alpha})^{1/2}  \tilde c_{\alpha,\text{in}}(t)
\end{equation}
that can be solved by Fourier transform:
\beq
	 \tilde d(\omega) = 
	 -i \chi(\omega) \sum_\alpha e^{i\phi_\alpha} ({\Gamma_\alpha}/{\pi\ \rho_\alpha})^{1/2} 
	 \ \tilde c_{\alpha,\text{in}}(\omega);
\eeq
where we have defined $\chi(\omega) = \left(i\omega + \Gamma \right)^{-1}$. 

In order to be able to compute the correlation functions we need  to specify the 
averages of the $c$-operators for $t=0$.
We assume that the Fermions are in thermal equilibrium for $t=0$:
\begin{eqnarray}
	\langle c^\dag_{\alpha,k}(t) c^{\phantom \dag}_{\beta,k'}(0) \rangle &=& \delta_{\alpha,\beta} \delta_{k,k'} 
	e^{i \epsilon_{\alpha k} t} f_F(\epsilon_{\alpha k}-\mu_\alpha) 
	\\
	\langle c^{\phantom \dag}_{\alpha,k}(t) c^\dag_{\beta,k'}(0) \rangle &=& \delta_{\alpha,\beta} \delta_{k,k'} 
	e^{-i \epsilon_{\alpha k} t} [1-f_F(\epsilon_{\alpha k}-\mu_\alpha)]
\end{eqnarray}
with vanishing $\langle cc \rangle$ and $\langle c^\dag c^\dag$ and where we have defined the Fermi distribution 
function $f_F(\epsilon)=1/(1+e^{\epsilon/T})$. 
One can then obtain the form of the correlation function for $d$:
\beq
	 \left\langle d^\dagger(\tau) \ d(0) \right\rangle_\omega 
	 = 
	 2\left|\chi(\omega - \epsilon_d)\right|^2 \sum_\alpha \Gamma_\alpha \ f_\alpha (\omega)
	 \label{corrdd}
\eeq
where $f_\alpha(\omega)=f_F(\omega-\mu_\alpha)$.

At equal time ($t=0$) \refe{corrdd} gives the average charge on the dot:
\begin{equation}
\label{eq:nbar}
        \langle n \rangle = 2 \int_{-\infty}^{+\infty}{\frac{d\omega}{2\pi} 
        \frac{\Gamma_L f_L(\omega) + \Gamma_R f_R(\omega)}{\Gamma^2 + (\omega - \epsilon_d)^2}} ,
\end{equation}
that leads to \refe{eq:FAD_electronicN} in the main text.

We consider now $S_{nn}(t)$. By using Wick theorem we can write it as a product of two correlation functions
of the form (\ref{corrdd}): $S_{nn}(t)=\langle d^\dag(t)d(0)\rangle \langle d(t)d^\dag(0)\rangle $.
The Fourier transform gives then:
\begin{equation} 
\label{eq:S_nn_approx}
        S_{nn}(\omega) = \sum_{\alpha, \beta} \int_{-\infty}^{+\infty} \frac{d\omega'}{2\pi} \frac{2\Gamma_\alpha f_\alpha (\omega')}{{\Gamma^2 + (\omega' - \epsilon_d)^2}} \cdot \frac{2\Gamma_\beta \left[1 - f_\beta (\omega'-\omega)\right]}{{\Gamma^2 + (\omega' - \omega - \epsilon_d)^2}}
\end{equation}
from which the expressions \refe{eq:FAD_electronicD} and \refe{eq:FAD_electronicA} can be derived.


\bibliographystyle{apsrev4-1}
\bibliography{FabioBib}

\end{document}